\global\def\draftcontrol{0}

   \def\versionno{From Courant to actions}

\catcode`\@=11

\expandafter\ifx\csname draftcontrol\endcsname\relax\global\def\draftcontrol{0}
\fi

{\count255=\time\divide\count255 by 60
\xdef\hourmin{\number\count255}
\multiply\count255 by-60\advance\count255 by\time
\xdef\hourmin{\hourmin:\ifnum\count255<10 0\fi\the\count255}}
\def\draftdate{\number\month/\number\day/\number\year\ \ \ \hourmin }

\newcommand\makepapertitle{\par
  \begingroup
    \renewcommand\thefootnote{\@fnsymbol\c@footnote}%
    \def\@makefnmark{\rlap{\@textsuperscript{\normalfont\@thefnmark}}}%
    \long\def\@makefntext##1{\parindent 1em\noindent
            \hb@xt@1.8em{%
                \hss\@textsuperscript{\normalfont\@thefnmark}}##1}%
     \newpage
     \global\@topnum\z@   
     \@makepapertitle
     \thispagestyle{empty}\@thanks
  \endgroup
  \setcounter{footnote}{0}%
  \global\let\thanks\relax
  \global\let\makepapertitle\relax
  \global\let\@makepapertitle\relax
  \global\let\@thanks\@empty
  \global\let\@author\@empty
  \global\let\@date\@empty
  \global\let\@title\@empty
  \global\let\title\relax
  \global\let\author\relax
  \global\let\date\relax
  \global\let\and\relax
  \def\version{\let\version\@version\@gobble}
}
\def\@makepapertitle{%
  \newpage
   \ifnum\draftcontrol=1 {}
   \version\versionno
   \vskip 3em%
   \else
   \hfill\hbox to 3cm {\parbox{4cm}{\@pubnum}\hss}%
   \vskip 3em%
   \fi
   \begin{center}%
   \let \footnote \thanks
     {\LARGE {\@title}}%
     \vskip 1.5em%
     {\normalsize
       \lineskip .5em%
       \begin{tabular}[t]{c}%
         \@author
       \end{tabular}\par}%
     \vskip 1.5em%
     {\@bstract}%
     \end{center}%
     \vskip 1.5em
     \@date%
   \par
}

\gdef\@pubnum{}
\def\pubnum#1{%
  \gdef\@pubnum{#1}}

\gdef\@bstract{}
\def\Abstract#1{%
  \gdef\@bstract{%
   \parbox{\textwidth-0pc}{%
   \centerline{\bf Abstract}\penalty1000%
\noindent
\renewcommand\baselinestretch{1.0}%
{#1}}}
}

\def\ps@paper{\let\@mkboth\@gobbletwo%
     \ifnum\draftcontrol=1
        \def\@oddfoot{\hbox to \textwidth{\tiny \versionno \hfil\tiny\draftdate}%
        \hskip -\textwidth \hbox to \textwidth{\hfil\rm\thepage\hfil}}%
     \else\def\@oddfoot{\hbox to \textwidth{\hfil\rm\thepage\hfil}}
     \fi
     \let\@evenfoot\@oddfoot
}

\def\@version#1{\ifnum\draftcontrol=1
\typeout{}\typeout{#1}\typeout{}
\vskip3mm\centerline{\hbox{\fbox{\normalsize{\tt DRAFT -- #1 -- }
                   {\draftdate}}}}\vskip3mm
\fi}
\let\version\@version
\long\def\eqlabel#1{\ifnum\draftcontrol=1
                    \tag@false  
                    \tag*{(\theequation) \hbox to -0.2cm{\hspace{0cm}\small{#1}\hss}}
                    \refstepcounter{equation}
                    \edef\@currentlabel{\theequation}
                    \ltx@label{#1}          
                    \else
                    \label{#1}
                    \fi
                    }
\let\st@bibitem\@bibitem
\let\st@lbibitem\@lbibitem
\ifnum\draftcontrol=1
  \def\@bibitem#1{%
    \st@bibitem{#1}\a@@label{#1}\ignorespaces}
  \def\@lbibitem[#1]#2{%
    \st@lbibitem[#1]{#2}\a@@label{#2}\ignorespaces}
  \def\a@@label#1{%
    \gdef\a@lab{\smash{\normalfont\small#1}}
    \ifvmode
      \if@inlabel
        \global\setbox\@labels\hbox{%
          \llap{\a@lab\let\a@lab\relax
                \kern\@totalleftmargin\kern\marginparsep}%
          \box\@labels}%
      \fi
    \fi}
\fi

\documentclass[12pt,letterpaper]{article}

\usepackage{amsmath,amssymb,array,calc,rotating,epsfig,psfrag}
\usepackage[nosort]{cite}
\usepackage{graphicx}
\usepackage{hyperref}

\ifnum\draftcontrol=1
\fi

\renewcommand\baselinestretch{1.25}
\renewcommand\section{\@startsection {section}{1}{\z@}%
                                   {-3.5ex \@plus -1ex \@minus -.2ex}%
                                   {2.3ex \@plus.2ex}%
                                   {\normalfont\large\bfseries}}
\renewcommand\subsection{\@startsection{subsection}{2}{\z@}%
                                   {-3.25ex\@plus -1ex \@minus -.2ex}%
                                   {1.5ex \@plus .2ex}%
                                   {\normalfont\normalsize\bfseries}}
\renewcommand\subsubsection{\@startsection{subsubsection}{3}{\z@}%
                                   {-3.25ex\@plus -1ex \@minus -.2ex}%
                                   {1.5ex \@plus .2ex}%
                                   {\normalfont\normalsize\it}}
\renewcommand\paragraph{\@startsection{paragraph}{4}{\z@}%
                                   {-3.25ex\@plus -1ex \@minus -.2ex}%
                                   {1.5ex \@plus .2ex}%
                                   {\normalfont\normalsize\bf}}

\def\revise#1       {\raisebox{-0em}{\rule{3pt}{1em}}%
                     \marginpar{\raisebox{.5em}{\vrule width3pt\
                     \vrule width0pt height 0pt depth0.5em
                     \hbox to 0cm{\hspace{0cm}{%
                     \parbox[t]{4em}{\raggedright\footnotesize{#1}}}\hss}}}}

\def\calg         {{\cal G}}

\def\call         {{\cal L}}

\def\del          {\partial}

\def\tr           {\mathop{\rm Tr}}

\def\half{{\frac12}}

\def\sqr#1#2{{\vcenter{\vbox{\hrule height.#2pt
 \hbox{\vrule width.#2pt height#1pt \kern#1pt
 \vrule width.#2pt}\hrule height.#2pt}}}}

\def\a{\alpha}
\def\b{\beta}
\def\r{\rho}

\def\O{\Omega}

\def\o{\omega}

\def\m{\mu}
\def\g{\gamma}
\def\l{\lambda}
\def\n{\nu}
\def\bn{\bar{\nu}}
\def\bm{\bar{\mu}}

\catcode`\@=12

\begin{document}




\newcommand{\be}{\begin{equation}}
\newcommand{\ee}{\end{equation}}
\newcommand{\beq}{\begin{equation}}
\newcommand{\eeq}{\end{equation}}
\newcommand{\ba}{\begin{eqnarray}}
\newcommand{\ea}{\end{eqnarray}}
\newcommand{\nn}{\nonumber}

\def\vol{\bf vol}
\def\Vol{\bf Vol}
\def\del{{\partial}}
\def\vev#1{\left\langle #1 \right\rangle}
\def\cn{{\cal N}}
\def\co{{\cal O}}
\def\IC{{\mathbb C}}
\def\IR{{\mathbb R}}
\def\IZ{{\mathbb Z}}
\def\RP{{\bf RP}}
\def\CP{{\bf CP}}
\def\Poincare{{Poincar\'e }}
\def\tr{{\rm tr}}
\def\tp{{\tilde \Phi}}
\def\Y{{\bf Y}}
\def\te{\theta}
\def\bX{\bf{X}}

\def\TL{\hfil$\displaystyle{##}$}
\def\TR{$\displaystyle{{}##}$\hfil}
\def\TC{\hfil$\displaystyle{##}$\hfil}
\def\TT{\hbox{##}}
\def\HLINE{\noalign{\vskip1\jot}\hline\noalign{\vskip1\jot}} 
\def\seqalign#1#2{\vcenter{\openup1\jot
  \halign{\strut #1\cr #2 \cr}}}
\def\lbldef#1#2{\expandafter\gdef\csname #1\endcsname {#2}}
\def\eqn#1#2{\lbldef{#1}{(\ref{#1})}%
\begin{equation} #2 \label{#1} \end{equation}}
\def\eqalign#1{\vcenter{\openup1\jot   }}
\def\eno#1{(\ref{#1})}
\def\href#1#2{#2}
\def\half{{1 \over 2}}

\def\ads{{\it AdS}}
\def\adsp{{\it AdS}$_{p+2}$}
\def\cft{{\it CFT}}

\newcommand{\ber}{\begin{eqnarray}}
\newcommand{\eer}{\end{eqnarray}}

\newcommand{\bea}{\begin{eqnarray}}
\newcommand{\eea}{\end{eqnarray}}

\newcommand{\beqar}{\begin{eqnarray}}
\newcommand{\cN}{{\cal N}}
\newcommand{\cO}{{\cal O}}
\newcommand{\cA}{{\cal A}}
\newcommand{\cT}{{\cal T}}
\newcommand{\cF}{{\cal F}}
\newcommand{\cC}{{\cal C}}
\newcommand{\cR}{{\cal R}}
\newcommand{\cW}{{\cal W}}
\newcommand{\eeqar}{\end{eqnarray}}
\newcommand{\lm}{\lambda}\newcommand{\Lm}{\Lambda}
\newcommand{\eps}{\epsilon}


\newcommand{\nonu}{\nonumber}
\newcommand{\oh}{\displaystyle{\frac{1}{2}}}
\newcommand{\dsl}
  {\kern.06em\hbox{\raise.15ex\hbox{$/$}\kern-.56em\hbox{$\partial$}}}
\newcommand{\as}{\not\!\! A}
\newcommand{\ps}{\not\! p}
\newcommand{\ks}{\not\! k}
\newcommand{\D}{{\cal{D}}}
\newcommand{\dv}{d^2x}
\newcommand{\Z}{{\cal Z}}
\newcommand{\N}{{\cal N}}
\newcommand{\Dsl}{\not\!\! D}
\newcommand{\Bsl}{\not\!\! B}
\newcommand{\Psl}{\not\!\! P}
\newcommand{\eeqarr}{\end{eqnarray}}
\newcommand{\ZZ}{{\rm \kern 0.275em Z \kern -0.92em Z}\;}

\def\s{\sigma}
\def\a{\alpha}
\def\b{\beta}
\def\r{\rho}
\def\d{\delta}
\def\g{\gamma}
\def\G{\Gamma}
\def\ep{\epsilon}
\makeatletter \@addtoreset{equation}{section} \makeatother
\renewcommand{\theequation}{\thesection.\arabic{equation}}

\def\be{\begin{equation}}
\def\ee{\end{equation}}
\def\bea{\begin{eqnarray}}
\def\eea{\end{eqnarray}}
\def\m{\mu}
\def\n{\nu}
\def\g{\gamma}
\def\p{\phi}
\def\L{\Lambda}
\def \W{{\cal W}}
\def\bn{\bar{\nu}}
\def\bm{\bar{\mu}}
\def\bw{\bar{w}}
\def\ba{\bar{\alpha}}
\def\bb{\bar{\beta}}

\begin{titlepage}

\version\versionno

\leftline{\tt hep-th/0610021}

\vskip -.8cm

\rightline{\small{\tt MCTP-06-25}}

\vskip 1.7 cm

\centerline{\bf \Large  A Geometric Action for the Courant Bracket}

\vskip 1.5cm {\large } 

\centerline{\large Xiaolong Liu${}^1$, Leopoldo A. Pando Zayas${}^2$,}

\vskip 0.5cm 
\centerline{\large  Vincent G. J. Rodgers${}^1$ and Leo Rodriguez${}^1$}

\vskip 1cm
\centerline{\it ${}^1$ Department of Physics and Astronomy}
\centerline{ \it  The University of Iowa}
\centerline{\it Iowa City, IA 52242}

\vskip .5cm
\centerline{\it ${}^2$ Michigan Center for Theoretical
Physics}
\centerline{ \it Randall Laboratory of Physics, The University of
Michigan}
\centerline{\it Ann Arbor, MI 48109-1040}

\vspace{1cm}

\begin{abstract}
An  important operation in generalized complex geometry is the Courant bracket which extends  
the Lie bracket that acts only on vectors to a pair given by 
a vector and a p-form. 
We explore the possibility of promoting the elements 
of the Courant bracket to physical fields by constructing a geometric action 
based on the Kirillov-Kostant symplectic form. For the $p=0$ forms, the action generalizes Polyakov's two-dimensional quantum gravity when viewed as the geometric action for the Virasoro algebra. We show that the 
geometric action arising from the centrally extended Courant bracket for the vector and zero form  pair is similar to the geometric action obtained from the semi-direct product of the Virasoro algebra with a $U(1)$ affine Kac-Moody algebra. For arbitrary $p$ restricted to a Dirac structure, we derived the geometric action and exhibit  generalizations for almost complex structures built on the Kirillov-Kostant symplectic form. In the case of $p+1$ dimensional submanifolds, we also discuss a generalization of a K\"ahler structure on the orbits of  $T^*\oplus \wedge^p T$. 
\end{abstract}



\end{titlepage}



\section{Introduction }
Understanding fluxes in string theory is one of the most active recent directions. 
Two immediate applications are found in the AdS/CFT correspondence and in moduli stabilization which plays 
a central role in 
phenomenological string model building. One of the best understood situations pertains to the 
B-field. Recently, there has been a successful cross fertilization between the mathematical and 
physics communities on this topic. For a review see \cite{Grana:2005jc}. In particular, generalized complex geometry has been shown to 
be relevant for various aspects of string theory with fluxes. 

Generalized complex geometry has been introduced by Hitchin  as a form of constructing 
differential geometry with a B-field \cite{hitchin, hitchinkitp}. More mathematically, it is a generalization 
of complex geometry, which includes the tangent bundle 
of a manifold $T$, to give $T\oplus {T}^*$, that is, the sum of 
the tangent and the cotangent 
bundle. A more complete account of generalized complex geometry can be found in 
\cite{gualtieri}. 
The interplay of this concept with the physics of supersymmetric
sigma-models is still being explored \cite{Lindstrom:2004iw}  but has already clarified various interesting properties of 
background with fluxes (for reviews see \cite{gcgphysics}). 
An example of generalized complex geometry is bi-Hermitian geometry which 
was already known in physics more than twenty years ago due to the work of Gates, Hull and Rocek \cite{ghr}.

An important structure in generalized complex geometry is the Courant bracket. In particular, integrability of the 
almost complex structures in generalized complex geometry is determined by the Courant bracket (see, for 
example,  section 4.3 of \cite{gualtieri}). The Courant bracket can be viewed as an extension of the 
Lie bracket of vector fields by acting on pairs 
given by a vector and a $p$-form \cite{courant}.   Part of what we will show is that for the particular case of $p=0$, the Courant bracket furnishes 
a generalization of the Virasoro algebra. This becomes manifest when one writes the geometric action which arises from integrating a symplectic two-form, $\Omega_{\tilde X}$, on the orbit of an element, say ${\tilde X}$, dual to $T\oplus \wedge^0T^*$. This implies a natural generalization of the 2D Polyakov action.  

Furthermore, the $p=1$ case, corresponding to $T\oplus T^*$, has been heavily exploited in determining generalized complex structures \cite{hitchin,hitchinkitp,gualtieri} that are related to $B$-fields.  This is done on a Dirac subbundle $E\in T\oplus T^*$. One of the constructions of a generalized complex structures relies on a symplectic two-form $\omega$ on the vector space $T$ so that relative to the the space $T \oplus T^*$ one may write 
\be {\cal J}_\omega = \left(
\begin{array}[h]{cc}
0 & -\omega^{-1}\\
\omega & 0	
\end{array}
\right). \label{1.1}
\ee
We are interested in exploring a somewhat different generalization of this which exploits the symplectic two-form $\Omega_{\tilde X}$ that lives on the orbit of ${\tilde X}\in E^*$.   This is the coadjoint orbit construction of the geometric action.  This will give a family of symplectic structures parameterized by the fields ${\tilde X}$.  In future work we hope to show that ${\tilde X}$ can be made  dynamical which imparts a variational principle to the space of symplectic structures.  Since $\Omega_{\tilde X}$ is a natural two-form on $E^*$ the hope is to relate this to generalized complex structures with the aim being whether such an approach can bring a resolution to at least part of the landscape problem in string theory.  

Finally one can use the above mentioned coadjoint orbit construction for the case when $p>1$.  This allows us to extend the notion of generalized complex structures to sections on $T\oplus \wedge^p T^*$.  Such bundles may be relevant for charged $D$-branes.  Thus we explore the rich structure between geometric actions which  are closely tied to sigma models and the complex structures which are tied to the symplectic geometry via the Courant bracket.  

Thus, our interests in constructing geometric actions for the Courant bracket 
are in the generalization of the Virasoro algebra for the case when $p=0$, developing a theory in which there is a natural potential for complex structures that come from the $p=1$ sector, and extending the differential geometry  to the $p>1$ sector by looking at $p$-extended generalized complex structures and their associated K\"ahler structure. All in all, we hope to clarify the role of the Courant bracket as a central element in the physics of generalizations of Calabi-Yau manifolds and $D$-brane physics.  

Our approach is to study the Courant bracket as an algebra and to construct its dual.  Part of this work will focus on the Dirac subbundle of the Courant bracket since this sector can be treated as a Lie algebra.  Using the methods of coadjoint orbits \cite{Kirillov,KK,Kirillov2} we construct an invariant action that is akin to the WZW model, 2D Polyakov actions and other sigma models by using the natural symplectic structure on each orbit.  We show an extension of Eq.[\ref{1.1}] for sections of the space $(T\oplus {\wedge}^p T^*)\otimes(T^*\oplus {\wedge}^p T)$ as an example of a $p$-extended generalized complex structure.  We further show that one may define a $p$-extended generalized   K\"ahler metric on each orbit for $p\ge 1$ when restricted to $p+1$ dimensional submanifolds.  In the particular case of $p=0$, we will construct the geometric action  for the Courant bracket and show that its central extension is a generalization of the Liouville action for the 
Virasoro algebra.

The paper is organized as follows. Section \ref{gcg} contains a brief review of generalized complex geometry 
and introduces the Courant bracket. Then in section \ref{DualCourant} we discuss the dual representation for the Courant bracket with arbitrary $p$ and show how the dual representation transforms under the action of the Courant bracket.  Section \ref{sectionaction} reviews in detail the construction of geometric actions 
and presents Polyakov's two-dimensional quantum gravity as the geometric action for the Virasoro algebra. In section \ref{explicitzero} we explicitly construct the action that 
results from the Courant bracket in the case of $p=0$. We also consider the central extension of this case and find that it is, in many aspects, similar to the action obtained from the semi direct product of the Virasoro algebra with a Kac-Moody affine 
algebra with group $U(1)$.  Section \ref{explicitall} discusses the general case $p\ne 0$. We find a geometric action for a Dirac subbundle of elements in $T\oplus {\wedge}^p T^*$.  In section \ref{6} we further discuss the $p$-extended generalized complex structures for this Dirac subbundle and present a $p$-extended K\"ahler structure on $p+1$ dimensional submanifolds. We present our conclusions and direction for future research in section \ref{conclusions}.

\section{Review of the Courant Bracket}\label{gcg}

Motivated by the prominent role that the $B$-field 
plays in string theory Hitchin \cite{hitchin, hitchinkitp} initiated a program to study differential geometry by considering structures 
in $T\oplus T^*$ rather than the standard tangent bundle $T$. An element of this generalized 
bundle is of the form $(X,\xi)$ where $X$ is a vector and $\xi$ a one-form. This bundle comes with a 
natural indefinite metric via the interior product of a one-form and a 
vector: $\langle (X,\xi), (X,\xi) \rangle= -i_X\xi$. 
The $B$-field appears naturally in this picture since a two-from  naturally introduces an automorphism of 
$T\oplus T^*$, i.e. $B: (X,\xi) \mapsto (X,\xi+i_X B)$. Thus, in this picture the $B$-field  generates 
isometries of the natural metric discussed above since $i_X i_X B=0$.

In principle, there is no canonical Lie bracket 
for this bundle, one can however introduce the Courant bracket which has a number of interesting and natural 
properties. The Courant bracket is a generalization of the Lie bracket on sections
of the tangent bundle $T$ to sections of the bundle $T\oplus \wedge^p T^*$. 
Let $X$ and $Y$ be vector fields and $\xi$ and $\eta$ represent 
$p$-forms.  Call ${\cal X}=(X,\xi)$ and ${\cal Y}=(Y,\eta)$. The Courant bracket is defined  on  ${\cal T}=(T,\b)\in
\mathbb{C}^\infty(T\oplus {\wedge}^pT^*)$ as \cite{hitchin, hitchinkitp,gualtieri}
\be
\eqlabel{courant}
[{\cal X, Y}]=[(X,\xi),(Y,\eta)]=\left([X,Y], \call_X\eta-\call_Y\xi-\frac12 \, d\big[i_{X}\eta-i_Y\xi\big]\right).
\ee
It will be useful for us to write the pair $(X,\xi)$ as a rank one contravariant tensor field and an anti-symmetric $p$-tensor viz, 
\be
(X,\xi)\rightarrow (X^b, \xi_{b_1 b_2 \cdots b_p}).
\ee
Then Eq.[\ref{courant}] has the explicit realization as 
\be
[(X^a,\xi_{\,a_1 a_2 \cdots a_p}),(Y^b,\eta_{\,b_1 b_2 \cdots b_p})] = (Z^c,\zeta_{\,c_1 c_2 \cdots c_p}),
\ee
where
\be
Z^c= X^a \partial_a Y^c - Y^a \partial_a X^c,
\ee 
and
\be 
\zeta_{\,c_1 \cdots c_p} =  {\cal L}_X \eta_{\,c_1  \cdots c_p} - {\cal L}_Y  \xi_{\,c_1  \cdots c_p} -\frac{1}{2} \partial_{\,[c_1} X^a \eta_{ |a| c_2 \cdots c_p ]} +\frac{1}{2} \partial_{\,[c_1} Y^a \xi_{|a| c_2 \cdots c_p]} .
\ee

As we just mentioned, an important property of the Courant bracket is that it allows 
non-trivial automorphisms defined by a closed $p+1$-form $\alpha\in
\mathbb{C}^\infty({\wedge}^{p+1}T^*)$:
\be
A(X,\xi)=(X, \xi+i_X\alpha)  \label{automorphism}
\ee
Using $[\call_X,i_Y]=i_{[X,Y]}$ one can easily check that
\be
A\left([(X,\xi),(Y,\eta)]\right)=[A(X,\xi), A(Y,\eta)].
\ee
In the case of $p=1$, as noted above, this automorphism is a natural place for 
introducing the $B$-field where in the above $\alpha =B$.

In general the Courant bracket does not satisfy the Jacobi identity.  Indeed the Jacobiator
\be
J({\cal X,Y,Z})={\cal 
[X,[Y,Z]]+ [Y,[Z,X]]+[Z,[X,Y]]}
\ee
does not vanish
due to the presence of terms of the type $i_X \eta$.
For the $p=1$ case, one circumvents the issue of non-associativity by focusing on the Dirac subbundle. 
{\it Definition:} A Dirac structure on $M$ is a subbundle $E\in T\oplus T^*$ such that  \cite{hitchin, hitchinkitp,gualtieri}: 
\begin{itemize}
\item  $E$ is maximally isotropic for natural metric, and
\item sections of $E$ are closed under the Courant bracket.
\end{itemize}
For our purposes for arbitrary $p$ it will be enough to think of the Dirac subbundle, $E_p$ as the subspace of all vector fields, $X$ and $p$-forms  $\beta$, such that $i_X \beta=0,  \forall  \ \{X, \beta\} \in E_p.$
One interesting case is $p=0$, in this case the Courant bracket does satisfy the Jacobi identity. It takes 
the form
\be
\eqlabel{courant00}
[(X,f),(Y,g)]=\left([X,Y], \call_X g-\call_Y f \right).
\ee

\section{The Dual Representation of the Courant Bracket}\label{DualCourant}
In order to discuss the dual of the Courant Bracket let us motivate the discussion by considering the dual of an algebra $\calg$ denoted by ${\calg}^*$. The action of the adjoint representation on itself 
is realized through the commutator, 
\be
X * Y =[X,Y], 
\ee 
where $X$ and $Y$ are in the adjoint representation.  We introduce a representation dual to the 
adjoint representation by constructing an invariant scalar 
\be
\langle \tilde X| Y\rangle=C({\tilde X}, Y),
\ee
where the $\tilde X$ is an element in the dual representation.  $C({\tilde X}, Y)$  represents a 
scalar with respect to the algebra so that for any $Z$ and $Y$ in the adjoint representation, and 
any $\tilde X$ in the dual representation
\be
Z * C({\tilde X}, Y)=0.
\ee

By using Leibnitz rule we can extract the action of the adjoint elements on the dual elements,
\be
Z*\langle \tilde X| Y\rangle=0 \Longrightarrow \langle Z* \tilde X| Y\rangle+ \langle \tilde X|  Z*Y\rangle=0,
\ee
and since the action of the adjoint representation on itself is known we conclude that 
\be
\langle Z* \tilde X| Y\rangle= - \langle \tilde X|  Z*Y\rangle = -\langle \tilde X|  [Z,Y]\rangle.
\ee
We define the coadjoint representation as the elements, say $\tilde X$, which are dual 
to the algebra so that the action of any element of the adjoint representation,  $Z*\tilde X$ is defined through
\be
\langle Z* \tilde X| Y\rangle = -\langle \tilde X|  [Z,Y]\rangle.
\ee

The group action of $G$ on $\tilde X$ is generated 
by the adjoint representation for those elements of the group that
are connected to the identity. 
Since the algebra generates the group $G$, one can use the algebra to make infinitesimal changes to the coadjoint element $\tilde X$.  For fixed $\tilde X$, the group action on $\tilde X$ defines the {\em coadjoint orbit} of $\tilde X$.  There are some group elements that leave $\tilde X$ invariant. 
The  isotropy algebra, $\cal H$ of $\tilde X$ is determined by the subalgebra of elements that send $\tilde X$ to zero.  In other words, $F$ is said to be in $\cal H$ if $F*\tilde X=0.$ $\cal H$ generates the {\em isotropy group} $H$.  The coadjoint orbit of $\tilde X$ is then characterized by the  coset space $G/H$.  The adjoint action of $\tilde X$ determines the tangent space on the orbit of $\tilde X$.  One may write that  $\delta_Z {\tilde X}=Z*\tilde X$ on the coadjoint orbit. If the algebra is Lie, then given any two coadjoint elements ${\tilde X}_1$ and ${\tilde X}_2$ on the orbit of ${\tilde X}$, one may define a symplectic two form by writing 
\be
\Omega_{\tilde X}({\tilde X}_1,{\tilde X}_2) = \langle {\tilde X} \mid [ F_1,F_2] \rangle, \label{Omega}
\ee 
where $\delta_{F_1} {\tilde X} = {\tilde X}_1$ and $\delta_{F_2} {\tilde X} = {\tilde X}_2$. 

Now the paradigm for constructing an action given an
algebra came from the seminal work of Kirillov \cite{Kirillov,KK,Kirillov2}, who recognized the important 
relation between symplectic structures and coadjoint orbits.
 The paradigm can be summarized for a symmetry group $G$ as:
\begin{itemize}
\item Write a pairing $< {\tilde X} \mid Y >$ between $Y$ an element of  $\cal{G}$ and ${\tilde X}$ an element of $\cal{G^*}$.
\item Using the adjoint representation, demand  that the pairing be invariant.  This allows us to define the coadjoint representation.  
\item The orbits of each element of the coadjoint representation, say ${\tilde X}$, corresponds to a symplectic manifolds with $G/H$ symmetry where $H$ is the subgroup that leaves ${\tilde X}$ invariant.  For $W$ and $Z$ elements of $\cal{G}$, the natural symplectic two-form on the orbit of ${\tilde X}$ is simply $\Omega_{\tilde X}(W,Z) \equiv < {\tilde X} \mid [W,Z] >.$
\item Use the natural symplectic structure on the orbits to build a \emph{Geometric Action},
$ S_{\tilde X}=\int \Omega_{\tilde X} $.  The fields in the geometric action correspond to elements of the group $G$ while ${\tilde X}$ is a background field that dictates the symmetry of the action.
\end{itemize}

It is symplectic since $\Omega_{\tilde X}({\tilde X}_1,{\tilde X}_2)$ is closed by virtue of the Jacobi identity of the 
algebra and nondegenerate \cite{Kirillov,KK,Kirillov2,witten}.  This is known in the literature as the Kostant-Kirillov form.
Using $\Omega_{\tilde X}({\tilde X}_1,{\tilde X}_2)$ the construction of the geometric action for the coadjoint orbit of ${\tilde X}$ is straightforward.  
To proceed with the construction of the geometric action,  one fixes the background fields ${\tilde X}$.  This chooses the symmetry of the action via $G/H$.  
Then by using the {\it group} action on ${\tilde X}$, integrates this symplectic two form over a two-dimensional manifold and recovers the geometric action.
\be
\eqlabel{action0}
S_{\tilde X}=\int_{\Sigma}\Omega_{\tilde X}.
\ee
This geometric action has inherited the $G/H$ symmetry of the orbit.  We come to the explicit construction 
of the action in sections \ref{explicitzero}. 
\begin{figure}[h]
    \centering
        \includegraphics[scale=.7,angle=0]{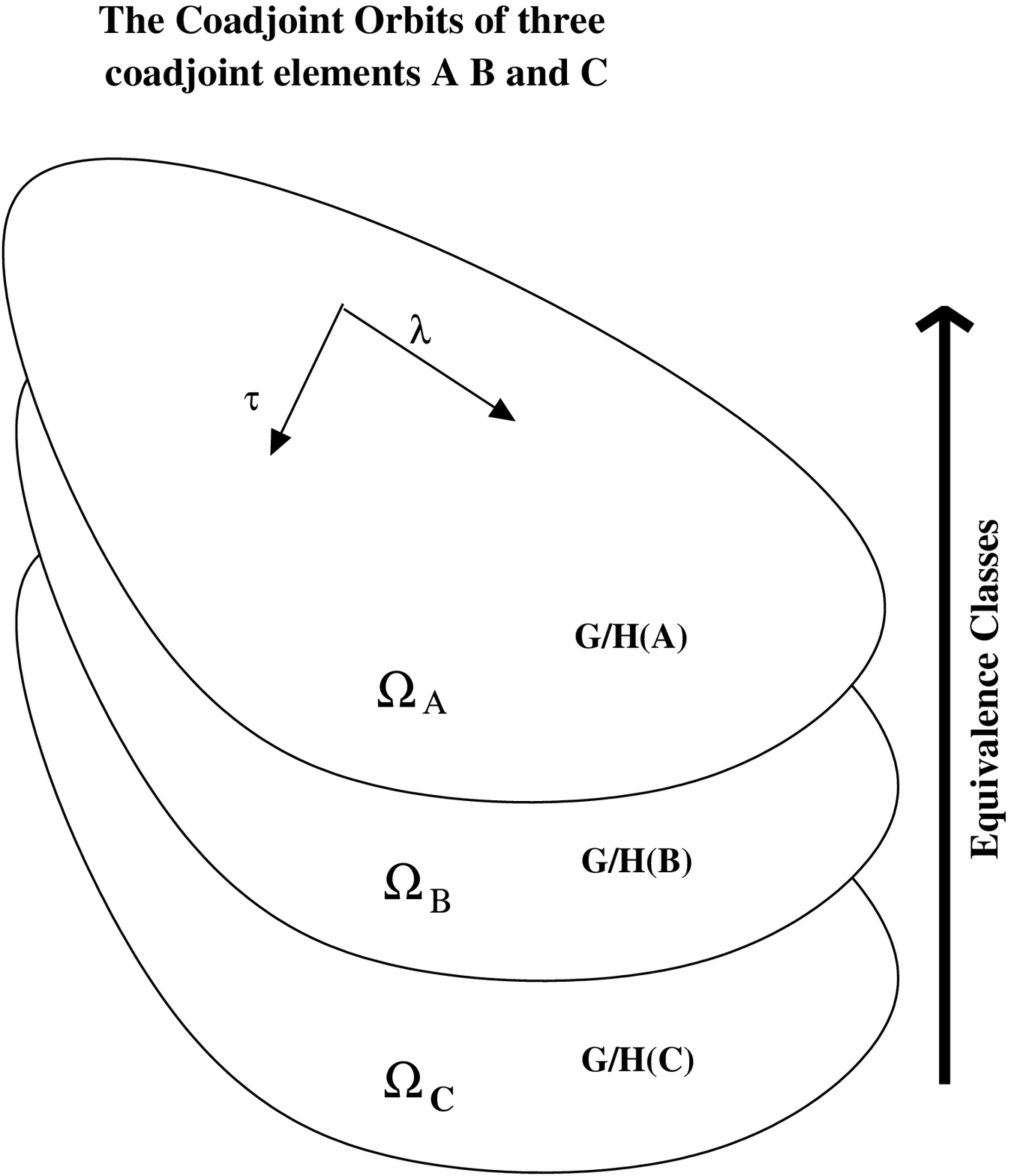}   
    \caption{Coadjoint Orbits }
    \label{fig:coadjoint_orbits}
\end{figure}

To implement the above procedure on $T\oplus {\wedge}^p T^*$ the bracket must be restricted to a subbundle where the Jacobi identity is satisfied.
It is for this reason that we restrict our study to a Dirac structure $E_p \in T\oplus {\wedge}^p T^*$. First we introduce the dual space of $E_p$, ${E^*_p}\in T^* \oplus {\wedge}^p T$,  The dual space elements  will be denoted by ${\tilde {\cal X}}=(\tilde X,\tilde \xi)$ which explicitly are the tensor densities of \textit{weight} $-1$, $(\tilde X_a , \tilde \xi^{a_1 \ldots a_p})$.  Then a suitable pairing over a volume $v$ between these spaces is:
\be
\langle (\tilde X,\tilde \xi)| (Y,\eta)\rangle =\a_1  \int dv \,\,\tilde X_a Y^a 
+ \a_2 \int dv \,\,\, \tilde \xi^{a_1 \ldots a_p} \,\eta_{a_1\ldots a_p}, \label{pairing}
\ee
where $\alpha_1$ and $\alpha_2$ are constants $dv$ is an infinitesimal volume element. 
Since we expect this pairing to be invariant under the action of the adjoint representation we have that 
\be
\langle (\tilde X,\tilde \xi)| (Z,\beta) *(Y,\eta)\rangle = -\langle (Z,\beta) * (\tilde X,\tilde \xi) | (Y,\eta)\rangle.
\ee
This relation defines the coadjoint action. One finds that the action of the adjoint element $(Z,\beta)$ on the coadjoint elements $\tilde{\cal X}=(\tilde X_a,\tilde \xi^{a_1 \ldots a_p})$ is 
\be
(Z,\beta)*( \tilde X_a,\tilde \xi^{a_1 \ldots a_p})= ( \d \tilde X_a,\d \tilde \xi^{a_1 \ldots a_p}),
\ee
where
\bea
\d \tilde X_a =&& -\a_1 \call_Z\,\tilde X_a
- \a_2 (\nabla_a \beta_{a_1\ldots a_p}) \tilde \xi^{ a_1\ldots a_p} 
- {(-1)^p}\a_2 \nabla_{[a_1}(\beta_{|a| a_2\ldots a_p]} \tilde \xi^{a_1\ldots a_p}) \nonumber \\
&& \ldots - {(-1)^p}\a_2 \nabla_{[a_p}(\tilde \xi^{a_1\ldots a_p} \beta_{ a_1 \ldots a_{p-1}]a})
-\frac{1}{2} \a_2 \nabla_{[a_1}(\tilde \xi^{a_1\ldots a_p})\beta_{|a|a_2\ldots a_p]} \nonumber \\
 \d \tilde \xi^{a_1 \ldots a_p} =&& -\a_2 \call_Z\,\tilde \xi^{a_1 \ldots a_p} 
+\frac12 \a_2 \,\,(\nabla_b\tilde \xi^{b\,[a_2\ldots a_p})  Z^{a_1]} . \label{deltaX}
\eea 
Note that the above expression is independent of the connection. 

From here the symplectic two-form on the orbit of ${\tilde {\cal Z}}$ is  straightforward. For  ${\tilde {\cal Z}}=({\tilde Z}_a, {\tilde \zeta}^{a_1 \cdots a_p})$ and the Courant elements ${\cal X}=(X^a, \xi_{a_1 \cdots a_p})$ and ${\cal Y}= (Y^a, \eta_{a_1 \cdots a_p})$, we write
\bea
&&\Omega_{\tilde {\cal Z}}({\cal X},{\cal Y}) = \int \int \int  dv \,d q_1\, d q_2 \, \omega_{I J}(q_1,q_2) {\cal X}(q_1)^I {\cal Y}(q_2)^J \nonumber\\
&=&\alpha_1 \int dv \, {\tilde Z}_a (X^c \nabla_c Y^a - Y^c \nabla_c X^a)\nonumber\\
&-& \alpha_2 \int dv \, {\tilde \zeta}^{a_1 \cdots a_p} \left( Y^a \nabla_a \xi_{a_1 \cdots a_p} + \xi_{a a_2 \cdots a_p} \nabla_{a_1} Y^a + \cdots + \xi_{a_1 \cdots a_{p-1} a} \nabla_{a_p} Y^a \right)\nonumber\\
&+& \alpha_2 \int dv \, {\tilde \zeta}^{a_1 \cdots a_p} \left( X^a \nabla_a \eta_{a_1 \cdots a_p} + \eta_{a a_2 \cdots a_p} \nabla_{a_1} X^a + \cdots + \eta_{a_1 \cdots a_{p-1} a} \nabla_{a_p} X^a \right). \nonumber\\
\label{twoform}
\eea

If we symbolically write the above as $\Omega_{I J} {\cal X}^I {\cal Y}^J$ we can identify the mapping of $\Omega_{I J}$ on each component and see that it maps $T\oplus {\wedge}^p T^*$ to $T^* \oplus {\wedge} T$.  
One finds that 
\be 
{(\Omega_{\tilde {\cal Z}})}_{I J} {\cal X}^I= (X^*_a, {\xi^*}^{\,\,\{b\}}) = \left( {\omega{(1)}}_{c\, a} X^c + {\omega{(2)}}_{a}^{\,\,\{b\}} \xi_{\{b\}}, \,\,{\omega{(3)}}^{\,\,\{b\}}_{a} X^a \right),\label{omegamap}
\ee
where $\omega(i)$ is associated with the $i^{th}$ summand in Eq.[\ref{twoform}]. 
Explicitly 
\bea
&&(\o(1)_{d a}\,X^d)[q_2] = \int \int dv \, dq_1\left({\tilde z}_a(X^c(q_1) \,\nabla_c \,\delta(q_2) - {\tilde Z}_c \,\delta(q_2)\,\nabla_a X^c(q_1))\right),\nonumber\\ 
&&(\o(2)_{a\,}^{\,\{d\}}\xi_{\{d\}})[q_2] =\nonumber\\
&& -\alpha_2 \int \int  dv \, dq_1 {\tilde \zeta}^{a_1 \cdots a_p} ( \delta(q_2) \,\nabla_a \xi(q_1)_{a_1 \cdots a_p} - \xi(q_1)_{a a_2 \cdots a_p} \nabla_{a_1} \,\delta(q_2) + \nonumber \\
&&\hskip100pt +\cdots + \xi(q_1)_{a_1 \cdots a_{p-1} a} \,\nabla_{a_p} \,\delta(q_2) ), \nonumber\\
&&(\o(3)_{a\,}^{\,\{d\}}X^{a})[q_2] =\nonumber\\
&&\alpha_2 \int \int dv \, dq_1 \, {\tilde \zeta}^{a_1 \cdots a_p} ( X(q_1)^a \delta^{d_1\cdots d_p}_{a_1 \cdots a_p} \,\nabla_a \,\delta(q_2)  + \delta(q_2)\,\delta^{d_1\cdots d_p}_{a a_2 \cdots a_p} \,\nabla_{a_1} \,X(q_1)^a  + \nonumber \\
&&\hskip100pt +\cdots + \delta(q_2)\,\delta^{d_1\cdots d_p}_{a_1 \cdots a_{p-1} a} \,\nabla_{a_p}\, X(q_1)^a ), 
\eea
where $\{b\}$ denotes ${b_1 \cdots b_p}$ and $\d^{\{a\}}_{\{b\}}\equiv \d^{a_1}_{b_1} \cdots \d^{a_p}_{b_p}$ .

 Similarly we can symbolically write the inverse on ${\cal X^*}_I = (X^*_a, {\xi^*}^{\{b\}})$  as 
\bea
&&{({\Omega^{-1}}_{\tilde {\cal Z}})}^{I J} {{\cal X}^*}_I = (X^a, {\xi}_{\,\{b\}}) \nonumber\\
 &&= \left( {{\omega{(3)}}^{-1}}^{\,d\,\,}_{\{b\}}{\xi^*}^{\,\,\{b\}},\,{{\omega(2)}^{-1 \,}}^{\,a}_{\,\,\{b\}}\left( {X^*}_{\,a} - {\omega{(1)}}_{\,\,a d}  {{\omega{(3)}}^{-1}}^{\,d\,\,}_{\{c\}} {\xi^*}^{\{c\}}\right)\right),\label{omegainversemap}\nonumber\\
\eea where $\{b\}$ denotes ${b_1 \cdots b_p}$.

For $p=1$ $\O_{\tilde Z}$ maps the space $T\oplus T^*$ back into itself.  
Indeed for this case one has 
\be \Omega_{\tilde{Z}} \left( \begin{array}[h]{c} X^a\\ \xi_a \end{array} \right)= \left(
\begin{array}[h]{cc}
\o(3) & 0\\
\omega(1) & \o(2)	
\end{array}
\right) \left( \begin{array}[h]{c} X^a\\ \xi_a \end{array} \right)  \label{p1omega}
\ee
and the inverse map
\be {\Omega}^{-1}_{\tilde{Z}}\left( \begin{array}[h]{c} X^a\\ \xi_a \end{array} \right)= \left(
\begin{array}[h]{cc}
{\o(3)}^{-1} & 0\\
-{\omega(2)}^{-1}\,\omega(1)\, {\omega(3)}^{-1}& {\o(2)}^{-1}	
\end{array}
\right) \left( \begin{array}[h]{c} X^a\\ \xi_a \end{array} \right). \label{p1omegainverse}
\ee
These will prove useful in section [\ref{6}] when we discuss the generalized complex structure for the $p=1$ case. 

Lastly, we would like the automorphism of Eq.[\ref{automorphism}] to generate an isotropy of the pairing, Eq.[\ref{pairing}].  This requirement induces a transformation on the dual space elements so that under the automorphism,
\be
A(\tilde X_a,\tilde \xi^{a_1 \ldots a_p})=(\tilde X_a + (-1)^p \a_{a \, a_1 \cdots a_p} \tilde \xi^{a_1 \ldots a_p},\, \tilde \xi^{a_1 \ldots a_p}). \label{dualautomorphism}
\ee
From here it is easy to see that the $\omega(1)$ component of $\Omega_{\tilde {\cal Z}}$ acts as an operator valued $B$ transformation for $p=1$.  Indeed  one observes that 
in Eqs.[\ref{omegamap},\ref{omegainversemap}],  ${\omega(1)}_{c\, a}$ acts as $\a_{c \,a}$ in both Eq.[\ref{automorphism}] and Eq.[\ref{dualautomorphism}] since it  is independently a closed two form. The  ${\tilde z}_a$ field governs the choice of this operator.

From here we  proceed to the construction the geometric action.  In the next section we give a brief review of the 2D Polyakov gravitational action as an example of a geometric action.  
\section{Review of the 2D Polyakov Geometric Action}\label{sectionaction}

Let us briefly review how the 2D gravity \`a la Polyakov arises as the geometric action of a particular 
orbit of the Virasoro algebra.  We view the
Virasoro group as the group of diffeomorphisms of a line or the circle ${\rm
Diff }S^1$. 

The Virasoro algebra is the centrally extended algebra of Lie derivatives in one dimension.  In  any dimension 
the algebra of Lie derivative with respect to $\xi$ on $\eta$ can be written as
\begin{equation}
{\cal L}_{\xi} \eta^a = \xi^b \partial_b \eta^a - \eta^b
\partial_b\xi^a = (\xi \circ \eta)^a,
\end{equation}
and satisfies,
\begin{equation}
 [{\cal L}_{\xi}, {\cal L}_{\eta}] = {\cal L}_{\xi \circ \eta}. 
\end{equation}
  
In {\em one} dimension, we can centrally extend this
algebra by including  a two cocycle which is coordinate 
invariant.  We write the commutation relations as 
\begin{equation}
 [({\cal L}_{\xi}, a), ({\cal L}_{\eta},b) ] = ({\cal L}_{\xi \circ \eta}, (\xi,\eta)).
\end{equation}
Here $(\xi,\eta)$ is called a two-cocycle and maps a pair of elements of the algebra into complex numbers. 
It satisfies the cocycle condition which are necessary conditions for the centrally extended algebra to satisfy the 
 Jacobi identity.  
We consider a two cocycle that depends on the one-dimensional metric $g_{a b}$, an arbitrary rank 
two tensor $D_{a b}$, and a constant, $c$:
\begin{equation}(\xi,\eta) = \frac{c}{2\pi} \int (\xi^a \nabla_a \nabla_b 
\nabla_c\eta^c)\,dx^b + \frac{1}{2\pi} \int (\xi^a D_{a b} \nabla_c \eta^c )\, dx^b -
(\xi\leftrightarrow \eta). \label{2cocycle}
\end{equation}
Here the index structure is left intact in order to show the invariance of the 
two cocycle. The one dimensional metric tensor  $g_{a b}$ is compatible with 
the covariant derivative operator $\nabla$.
Only the triple derivative term is special to the one dimensional construction 
as it will not satisfy the Jacobi identity in other than one dimension.  
The term containing $D_{a b}$ can exist as a two-cocycle for the algebra of Lie 
derivatives over any line integral in higher dimensions.  

Now taking advantage of the fact that we are in one dimension we write, 
\begin{equation}
(\xi,\eta) = \frac{c}{2\pi} \int (\xi \eta''' - \xi'''
  \eta)\,dx
+ \frac{1}{2\pi} \int (\xi \eta' - \xi' \eta )(D + c (\Gamma'-\Gamma^2/2)) \,dx.
\label{eq:4}
\end{equation} 
Here $\Gamma$ is the one dimensional Christoffel symbol $\Gamma^i_{\,\,j\,k}$ and  ${}^{\prime}$ denotes derivation with respect to the coordinate.   It is important to acknowledge that Eq.[\ref{2cocycle}] is not the most general form of a two-cocycle.  Indeed one can capture the connection independence of a two-cocycle by writing 
\begin{equation}
(\xi,\eta) = \frac{c}{2\pi} \int (\xi \eta''' - \xi'''
  \eta)\,dx
+ \frac{1}{2\pi} \int (\xi \eta' - \xi' \eta )(D + c {\cal R})\,dx,
\label{2cocycle2}
\end{equation}
where ${\cal R}$ is any projective connection that transforms as $\Gamma'-\Gamma^2/2$ under one dimensional coordinate transformations.  However for the sake of heuristics, we will use Eq.[\ref{eq:4}] in what follows. 
Our choice for the two cocycle, Eq.[\ref{eq:4}],  can be 
reduced to 
\begin{equation}(\xi,\eta) = \frac{c}{2\pi} \int (\xi \eta''' - \xi'''
  \eta)\,dx
+ \frac{1}{2\pi} \int (\xi \eta' - \xi' \eta )\,B \,\,dx, \label{pairing0}
\end{equation}
which depends on, $B=D + c\,(\G' - \frac 12 \G^2)$.  $B$ is called a {\em quadratic differential}.  This separation of the `triple derivative' term from the quadratic differential is due to the fact that the `triple derivative' term will separately satisfy the Jacobi identity.  Even though neither the `triple derivative' term nor the quadratic differential term are covariant expressions, the two-cocycle is still coordinate invariant.  Different choices of $B$ determine different symmetries.  

The Lie derivative of $B$ with respect to $\xi$ is 
\begin{equation}
\d B = 2 \xi' B + \xi B' +2c\, \xi'''. \label{liederivative}
\end{equation}
A suitable pairing between the centrally extended algebra 
element $(\xi^\beta, \alpha)$ and a coadjoint element $(B_{\rho \gamma}, b)$ may be written as 
\be 
\langle (B_{\rho \gamma},b) \mid (\xi^\beta, \alpha) \rangle = \frac{1}{2\pi} 
\int \xi^\beta B_{\beta \gamma} dx^\gamma + b\, \alpha.\label{pairing1}
\ee
Again, after suppressing the index structure, the centrally extended algebra element 
$$\textbf{F}=((\xi \eta' - \xi' \eta ),\int (\xi \eta''' - \xi''' \eta)\,dx )$$
 and the coadjoint element $\textbf{B}=( B, \frac{c}{2 \pi})$ can be paired to give 
the two-cocycle from Eq.[\ref{pairing0}];
\be
\langle \textbf{B} \mid \textbf{F} \rangle =
\langle ( B, \frac{c}{2 \pi}) \mid \left((\xi \eta' - \xi' \eta ),\int (\xi \eta''' - \xi''' \eta)\,dx \right) \rangle = (\xi,\eta). \label{fiducial}
\ee

It is worth remarking that the object $\G' - \frac 1{2} \G^2$ is precisely the Schwartzian derivative.  To see this consider a one dimensional coordinate transformation $x \rightarrow s(x)$.  Since 
\be
{\tilde g}_{\alpha \beta}(s) = \frac{\partial x^\mu }{ \partial s^\alpha}\frac{\partial x^\nu}{\partial s^\beta} g_{\mu \nu}(x)
\ee
and for $g_{\alpha \beta}(x)=1$, $\G= \frac{1}{ \partial_x s(x)} \partial^2_x s(x)$.  Then it follows that 
\be
\label{gammatransformation}
(\G' - \frac 1{2} \G^2)= -\frac{3}{2} \frac{ (\partial^2_x s(x))^2}{ (\partial_x s(x))^2}+ \frac{\partial^3_x s(x)}{\partial_x s(x)}=\{s(x),x\},
\ee
where $\{s(x),x\}$ is the Schwartzian derivative.  With this we know the finite transformation law 
of the coadjoint vector, $\textbf{B}=( B, \frac{c}{2\pi})$ with respect to the group element $s(x)$. The 
finite transformation law for $(B(x), c)$ is then
\begin{equation}
\textbf{B}^\prime(s')=\left( \left(\frac{B(x(s))}{(ds/dx)^2} + \frac{c \{ s, x \}}{2 \pi (ds/dx)^2} \right)_{s=s'}, \frac{c}{2 \pi} \right) \label{grouptransform}
\end{equation}

The above transformation determines the coadjoint orbit of $\textbf{B}$ since it determines all the coadjoint 
elements that can be reached by making a finite transformation on
${\textbf B}$ \cite{Kirillov,Kirillov2,witten}.  A coadjoint element, say ${\textbf A}$ that cannot be reached from the group action on ${\textbf B}$ has a separate orbit as in Figure[\ref{fig:coadjoint_orbits}].  
The collections of all orbits foliates the dual space of the algebra, ${\calg}^*$.  Each orbit admits a symplectic two form, yielding a Poisson bracket structure, which can be integrated over a suitable two manifold to produce a physical action. 
These actions are called {\em geometric actions} \cite{Rai,Alekseev:1988ce,Delius,Lano:1992tc,Lano:1994gx}. 
The symmetry of the action can also be extracted from Eq.[\ref{grouptransform}] and its infinitesimal 
counterpart Eq.[\ref{liederivative}] as the isotropy group $H({\textbf B})$ corresponds to those transformation that leave $\textbf{B}$ invariant or at the infinitesimal level, those vector fields $\xi$ in Eq.[\ref{liederivative}] that give zero variation.  Since the orbit of ${\textbf B}$ can be characterized by the coset $G/H({\textbf B})$ then the action corresponding to this orbit has $G/H({\textbf B})$ symmetry.

\begin{figure}[h]
    \centering
        \includegraphics[scale=.5]{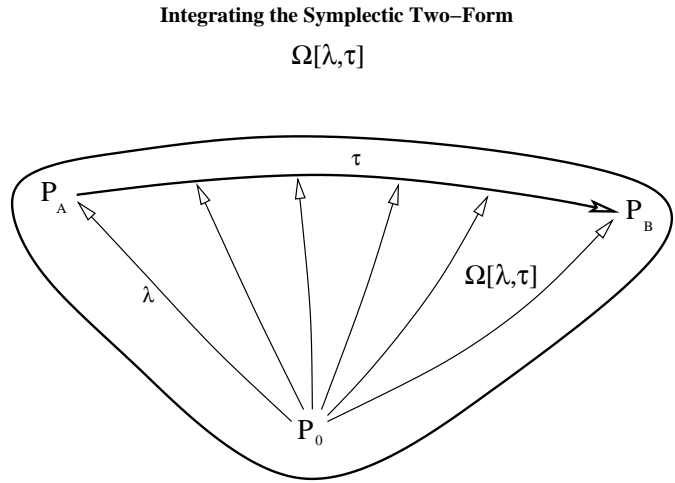}
    \caption{The Parameters $\lambda$ and $\tau$ sweep out a 2D submanifold.}
    \label{fig:symplectic}
\end{figure}

To construct the action we employ the techniques found in \cite{Balachandran}.  Consider any manifold M, endowed with a non-degenerate, closed two form $\Omega$, Figure[\ref{fig:symplectic}]. Then an action functional for any trajectory connecting two points, say $P_A$ and $P_B$, on M can be constructed by first choosing any arbitrary
but otherwise fixed point $P_0$.   Then for some trajectory parameterized by $\tau$ joining $P_A$ to $P_B$, one labels each of a one parameter family of paths joining $P_0$ to every point on the $\tau$ trajectory with a parameter $\lambda$ in such a way that when $\lambda$ equals 0 we are at the point $P_0$
and when $\lambda$ equals 1 we are at some point on the trajectory.  The family of paths
joining $P_0$ to the $\tau$ trajectory should be chosen so as to sweep out a two-dimensional
submanifold in M.  Then by integrating the two form $\Omega$ over this two dimensional surface,
we get the desired action,
\begin{equation}
S=\displaystyle \int_M \Omega
\end{equation}
where $M$ is the submanifold triangulated by the points $P_0,
P_A, {\rm and}\; P_B$.

In what follows we will consider $\tau$ to be an evolution parameter and $\lambda$ to represent 
an additional spatial coordinate that runs from $0 \le \lambda \le 1$. Thus,  our group 
elements may be written as $s(\lambda, \tau; x)$ corresponding to a two parameter family of 
diffeomorphism group elements. 
We will use boundary conditions so that $s(\lambda=0, \tau; x)=x$ 
and $s(\lambda=1, \tau; x)\equiv s(\tau, x)$. 

For the explicit construction of the action, we need to know the algebraic elements, i.e. the 
infinitesimal coordinate transformations along $\lambda$ and $\tau$. Then one may write that 
\be
S_B = \int d\lambda d\tau ~ \Omega \big( \textbf{B}^\prime_\lambda({ s(\lambda,\tau)}),
 \textbf{B}^\prime_\tau({s(\lambda,\tau)}\big)) =\nonumber \ee
\be \,\,\,\,\,\,\,
 \int d\lambda d\tau \bigl\langle \textbf{B}^\prime({s(\lambda,\tau)}) \bigr\vert
 \big[ {F}^\prime_\lambda({s(\lambda,\tau)}), {F}^\prime_\tau({s(\lambda,\tau)})
 \big] \bigr\rangle. \label{action1}
\ee
where here ${F}_\lambda({s(\lambda,\tau)})$ and 
${F}_\tau({s(\lambda,\tau)})$ are the elements of the algebra living in the adjoint representation that are
associated with infinitesimal transformations along the $\lambda$ and $\tau$ directions respectively. 
 With this we find that
\begin{eqnarray}
{F}^\prime_\lambda(s) &\equiv& s^{-1}{F}^s_\lambda s\cr
{F}^\prime_\tau(s) &\equiv & s^{-1}{F}^s_\lambda s \label{eqno(3)}
\end{eqnarray} and 
\begin{equation}
    s^{-1}{F}^s_\lambda s = \bigg( \frac{\partial_\lambda 
s(\lambda, \tau, x)}{ \partial_x s }, 0 \bigg), \qquad 
s^{-1}{F}^s_\tau s = \bigg( \frac{\partial_\tau s(\lambda, \tau, x)}{\partial_x s }, 0 \bigg).
\label{eqno(3a)}
\end{equation}
In the above, the quantity $s^{-1} {F}^s_{\tau}s$ represents the 
pull back of the adjoint vector to the $x$ coordinate system and ${\bf B}$ is also evaluated in the $x$ coordinate system.
From Eqs.[\ref{fiducial},\ref{action1}] one can show that the action can be written as
\be
S=\int d\lambda\,d\tau\biggl(\langle  B^\prime(s) \mid
 [s^{-1}F^{s}_\lambda {s} ,{s} ^{-1}
 F^{{s}}_\tau s]\rangle 
+  \frac{c}{2 \pi}({s}^{-1} F^{{s}}_{\lambda} {s},
{s}^{-1}F^{{s}}_{\tau}{s})\biggr)\ \label{action}
\ee
with
\be[s^{-1}F^{s}_\lambda {s} ,{s} ^{-1} F^{{s}}_\tau s] \equiv F_{\lambda} \left(
\frac{\partial}{\partial s'} F_{\tau} (s') \right) - \left(
\frac{\partial}{\partial s'} F_{\lambda} \right) F_{\tau} (s')\ee
and
\be({s}^{-1} F^{{s}}_{\lambda} {s}, {s}^{-1}F^{{s}}_{\tau}{s}) =  \int ds' \left( F_{\lambda} \frac{\partial ^3
F_{\tau}}{\partial s'^3} - F_{\tau}\frac{\partial^3}{\partial s'^3}
F_{\lambda}  \right). \label{secondterm}\ee
Explicitly in index notation this is 
\be
S=\int d\lambda\,d\tau\,ds^\gamma \,\bigl(B_{\alpha \beta}(s(x)) +\frac{c}{2\pi}(\partial_\mu \Gamma^\mu_{\alpha \beta} -\frac{1}{2} \Gamma^\mu_{\alpha \nu} \Gamma^\nu_{\beta \mu})\bigr)\frac{dx^\alpha}{ds^\rho}\frac{dx^\beta}{ds^\gamma}\bigl( \frac{\partial s^\xi}{\partial \tau} \frac{\partial^2 s^\rho}{\partial s^\xi \partial \lambda} - \frac{\partial s^\xi}{\partial \lambda} \frac{\partial^2 s^\rho}{\partial s^\xi \partial \tau}\bigr),
\ee
here the order of derivatives with respect to $s$ and the derivatives of  $\lambda$ and $\tau$ are 
important since they do not commute.
After suppressing the indices, contributions for that part of the action independent of $B_{\alpha \beta}(x)$ are:
\begin{eqnarray}
\frac{c}{2 \pi}\int d x d \lambda d \tau \{
\frac{3}{2} \frac{\partial s}{\partial \tau} 
\frac{\partial^2 s}{\partial \lambda \partial x} \frac{\partial^3 s/ \partial x^3}{(\partial s/ \partial x )^3}  - 3 \frac{\partial s}{\partial \tau}
\frac{\partial^2 s}{\partial \lambda \partial x} \frac{(\partial^2 s/ \partial x^2)^2}{(\partial s/ \partial x)^4} +\nonumber\\
\frac{3}{2} \frac{\partial^2 s}{\partial x^2} \frac{\partial^3 s}{\partial^2 x \partial \lambda
} \frac{1}{(\partial s/ \partial x)^3} \frac{\partial s}{\partial \tau}
- \frac{1}{2} \frac{1}{(\partial s/ \partial x)^2} \frac{\partial^4 s}{\partial x^3 \partial \lambda } \frac{\partial s}{\partial \tau}  \nonumber\\
-  (\lambda \leftrightarrow \tau).~~ \}
\end{eqnarray}
By including the $B(x)$ contribution we can write the action (up to total derivatives) as
\begin{equation}
S = \frac{c}{2 \pi} \int d x d \tau \left[
\frac{\partial^2_{x} s}{(\partial_{x}s)^2} \partial
_{\tau} \partial_{x} s - \frac{(\partial^2_{x}s)^2
(\partial_{\tau} s)}{(\partial_{x} s)^3} \right]
 - \int d x d \tau  B(x)
\frac{(\partial s/ \partial \tau)}{(\partial s/ \partial
x)}.
\end{equation}
If we change notation $\,x\rightarrow x_-$,$\tau \rightarrow x_+$, $s \rightarrow f,\,$ and $B\rightarrow 0$  then the action is identical to Polyakov's action \cite{pol}, viz
\begin{equation}
S = \frac{c}{2 \pi}\int d^2x \left[ \left( \partial^2_- f\right) \left(\partial_+ \partial_- f\right)
\left( \partial_- f \right)^{-2} - \left(\partial^2_- f\right)^2
(\partial_+ f) \left( \partial_-
f\right)^{-3} \right] - \int d^2x B \partial_+f/ \partial_-f .
\end{equation}

This is nothing but equation (5) in \cite{pol}. Alternatively, this is the light-cone gauge 
expression of equation (16) of \cite{polyakov} which results from integrating over the string embedding $X^\mu$ in the 
path integral with the standard Polyakov action $\int d^2 \xi \sqrt{g} g^{ab} \partial_a X^\mu \partial_b X_\mu$. The 
reader might perhaps be more familiar with the integration over string embedding $X^\mu$ with the 
world sheet metric in the conformal gauge, in which case the result is the Liouville action. 

\section{A geometric action for the Courant bracket}\label{explicitzero}
\subsection{Transformation Laws and Isotropy Equations for $p=0$}

A particularly interesting case is $p=0$ where in the pair under consideration $(X,f)$ the second entry is 
just a function $f$. This case was mentioned in 
\cite{hitchin} where it was noted that the Courant bracket can be understood as the usual Lie bracket 
on $S^1-$invariant vector fields of the form $X+f\frac{\partial}{\partial \theta}$ on $M\times S^1$. 
In this case the Courant algebra may be written as
\be
\eqlabel{courant0}
(X,f)*(Y,g)=[(X,f),(Y,g)]=\left([X,Y], \call_X g-\call_Y f\right).
\ee
First we define a pairing between the adjoint elements and their duals.   A suitable pairing  between the adjoint element $(Y,g)$ and the dual element $(\tilde X, \tilde f)$ for any number of dimensions may be written as
\be
\langle (\tilde X,\tilde f)| (Y,g)\rangle = \int dv \,\,\tilde X_a Y^a + \int dv \tilde f g.
\ee
The choice of scalar product is natural in the sense that it is a quadratic pairing and it is linear 
in the fields. This should be compared to the line integral pairing in Eq.[\ref{pairing1}] where the dual 
elements become a quadratic differential and where a central extension can exist. 
As an important remark we could have used this pairing for the Virasoro algebra where $dv$ is the one 
dimensional volume.  Instead of quadratic differentials the dual elements would have been covariant 
tensor densities of rank one and {\em weight} $-1$.  (In our notation the square root of determinant of the metric $\sqrt{g}$ has  weight $-1$.) Thus, the one dimensional pairing could also have been 
chosen to be 
\be 
\langle (B_{\rho},b) \mid (\xi^\beta, \alpha) \rangle_{\text{alternate}} = \frac{1}{2\pi}\int \xi^\beta 
B_{\beta} dx + b\, \alpha.\label{pairing2}
\ee
Recall that for a scalar density  of weight $n$, say $h$, that the covariant derivative of $h$ is 
given by 
$$\nabla_a h = \partial_a h + n h ~\Gamma^b_{~a b}.$$ This is consistent with $$ \nabla_a \sqrt{g} = \partial_a \sqrt{g} - \sqrt{g} ~\Gamma^b_{~a b}=0.$$  Similar modifications are true for the Lie derivative on densities. As an example, we have  
$$\delta g = g g^{a b} \delta g_{a b}$$ for any variation of a metric, so one finds that 
$${\cal L}_\eta g = g g^{a b} {\cal L}_\eta g_{a b}= - g g^{a b}(\nabla_{a} \eta_{b} + \nabla_{b} \eta_{a}) = - \eta^a \partial_a g - 2 g \partial_a \eta^a.$$
For a tensor density with weight $n$ the Lie derivative with respect to a vector field say, $\eta$, contains an extra summand $n ~\partial_a \eta^a$ from the Lie derivative of a tensor with the same rank but zero weight.  Thus the Lie derivative of $B^{\text{tensor}}_\rho$ which has rank one and weight $-1$, in the direction of $\eta^a$ is
\be 
\delta_\eta B^{\text{tensor}}_\alpha = - \eta^\lambda \partial_\lambda B^{\text{tensor}}_\alpha 
- B^{\text{tensor}}_\lambda \partial_\alpha \eta^\lambda -B^{\text{tensor}}_\alpha (\partial_\lambda \eta^\lambda).
\label{ambiguity1}
\ee
It is worth noting that in one dimension, that there is an ambiguity in tensor classifications.  Indeed any one dimensional tensor of rank $p$ and weight $w$ transforms like a tensor or rank any rank $q$ and that has weight $q-p+w$.  Thus in one dimension, Eq.[\ref{ambiguity1}] reduces to 
\be 
\delta_\eta B^{\text{tensor}} = - \eta (B^{\text{tensor}})^\prime - 2 B^{\text{tensor}} \eta^\prime \label{ambiguity2}.
\ee
This is the transformation law for a quadratic differential in one dimension.  

Now we need the transformation laws for the Courant coadjoint representation. Let 
${\cal F}=(X,h)$ and ${\cal G}=(Y,g)$ be two adjoint vectors. Since we demand that the pairing be invariant under the group transformation laws, the adjoint action on the pairing will give zero. By Leibnitz rule we can then find the transformation law for a coadjoint element ${\bf B}=(\tilde X,\tilde f)$ since
\be
{\cal F}*\langle {\bf B} \mid {\cal G} \rangle =0
\ee which is 
\be
(X,h)*\langle (\tilde X,\tilde f)| (Y,g)\rangle = 0.
\ee
Using Leibnitz rule,
\be
\langle (X,h)*(\tilde X,\tilde f)| (Y,g)\rangle + 
\langle (\tilde X,\tilde f)| (X,h)*(Y,g)\rangle =0,
\ee
leads to the transformation laws 
\be
\delta_{\cal F}{\bf B}=(X,h)*(\tilde X,\tilde f)=\left( \call_X \tilde X-
\tilde f d h, \, \call_X\tilde f\right). \label{transform1} 
\ee
Explicitly this is
\be \call_X \tilde X_a = - X^b \partial_b {\tilde X}_a - {\tilde X}_b \partial_a X^b - {\tilde X}_a \partial_b X^b\ee
and 
\be
\call_X \tilde f = -X^a \partial_a \tilde f - {\tilde f} \partial_a X^a
\ee
so that component-wise
\be
\delta_{\cal F}{\bf B}=\left( - X^b \partial_b {\tilde X}_a - {\tilde X}_b \partial_a X^b - {\tilde X}_a \partial_b X^b - \tilde f d_a h , -X^a \partial_a \tilde f - {\tilde f} \partial_a X^a \right). \label{transform2}
\ee
Eq.[\ref{transform2}] reveals that ${\tilde X}_a$ transforms under the coordinate transformations as a vector field density with weight $-1$.  (As 
we remarked in the beginning of this section, this is also a quadratic differential in one dimension due to the one dimensional ambiguity.) The 
element ${\tilde f}$ is a scalar density of weight $-1$.  The inhomogeneous contribution to the transformation 
law for ${\tilde X}$, i.e.  $- \tilde f d h$, is reminiscent of the transformation law for the quadratic 
differential in the presence of a gauge field \cite{Rai,Lano:1994gx,Branson:1996pe,Branson:1998bc}.  Later in section \ref{pzero} we will show that this is 
indeed the case in one dimension.  The field theory associated with the $p=0$ Courant bracket 
is equivalent to that of a $U(1)$ Kac-Moody algebra tensored with the Virasoro algebra.

The finite transformations associated with Eq.[\ref{transform2}] contains both a coordinate transformation and a shift are
\be
{\tilde X}_a(z) =\det\left({\partial y^c \over \partial z^e} \right)\left( {\partial y^b \over \partial z^a} X_b(y){+\tilde{f}{\partial y^b\over \partial z^a}{\partial h \over \partial y^b}}\right)
\ee
and
\be
{\tilde f}(z)= \det\left({\partial y^c \over \partial z^e} \right) f(y). 
\ee The $h$ transformations of Eq.[\ref{transform2}] adds an inhomogeneous contribution to ${\tilde X}_a$ that is akin to a U(1) gauge  transformations. The term $\tilde f d h$ will also transform as vector field with density $-1$ as we will see in Eq.[\ref{Bp0}].  
The importance of the finite coordinate transformation laws are that they allow us to go anywhere on 
the coadjoint orbit which is required to construct the invariant action.
Each orbit will inherit symmetries that are determined by those adjoint elements that leave 
the coadjoint element that defines the orbit invariant.  In general the group generated by 
the adjoint representation ${\cal G}$ and the subgroup that is generated by the isotropy 
algebra ${\cal H}$ characterize the coadjoint orbit through the coset ${\cal G}/{\cal H}$.

The symmetries of the coadjoint orbit associated with the $(\tilde X,\tilde f)$ can be determined from the isotropy algebra of $(\tilde X , \tilde f)$ which is defined by those elements of the adjoint elements,  which satisfy
\be
(X, h)*(\tilde X, \tilde f)=(0,0).
\ee
These elements form the subalgebra ${\cal H}$ mentioned above and lead to the symmetry relations
\be
\call_{X} \tilde X - \tilde f \,d h=0, \qquad 
\call_{X} \tilde f=0, \label{isotropy}
\ee
for $(X,f) \in {\cal H}$.
The above equations have the following geometrical interpretation. The
first equation corresponds to symmetries produced when certain gauge transformations 
can offset changes from certain coordinate transformations.  The relation can also be 
established by the family of gauge and coordinate transformations that separately 
leave ${\tilde X}$ and ${\tilde f}$ invariant.  These equations may also be interpreted 
as a geodetic equations, where the isotropy equation of ${\tilde f}$  requires that the 
quantity $\tilde f$ be conserved along the flow of $X$. Since $\tilde f$ is a density (volume), this 
equation denotes the Killing equation for volume preserving vector fields associated with $\tilde f$.

\subsection{The $p=0$ Geometric Action}

Let us now construct the invariant geometric action, $\Omega$. Guided by Eq.[\ref{action1}] we 
let ${ F}_\lambda({s^\mu(\lambda,\tau)},{h(\lambda,\tau)})$ and 
${ F}_\tau({s^\mu(\lambda,\tau)},{h(\lambda,\tau)})$  correspond to the generators along the 
two directions, $\lambda$ and $\tau$ respectively where here $s^\mu(x^\alpha; \lambda ,\tau)$ 
corresponds to a two parameter family of coordinates and ${h(\lambda,\tau)}$ is a two parameter 
family of functions of $x^\beta$.
\be
S_B = \int d\lambda d\tau ~ \Omega \big( \textbf{B}^\prime_\lambda({ s^\mu(\lambda,\tau)},{h(\lambda,\tau)}),
 \textbf{B}^\prime_\tau({s^\mu(\lambda,\tau)},{h(\lambda,\tau)}\big)) =\nonumber \ee
\be \,\,\,\,\,\,\,
 \int d\lambda d\tau \bigl\langle \textbf{B}^\prime({s^\mu(\lambda,\tau)},{h(\lambda,\tau)}) \bigr\vert
 \big[ {F}^\prime_\lambda({s^\mu(\lambda,\tau)},{h(\lambda,\tau)}), {F}^\prime_\tau({s^\mu(\lambda,\tau)},{h(\lambda,\tau)})
 \big] \bigr\rangle. \label{action2}
\ee
From Eq.[\ref{transform2}]  we find that
\be
\textbf{B}^\prime({s^\mu},{h}) = \bigg(\mid {\partial x^\mu \over \partial s^\nu} \mid (  {\tilde X}_\alpha {\partial x^\alpha \over \partial s^\rho} +   {\tilde f} {\partial x^\beta \over \partial s^\rho} {\partial h \over \partial x^\beta}), \,\, \mid {\partial x^\mu \over \partial s^\nu} \mid {\tilde f} \bigg)\label{Bp0} \ee
and
\be
\big[ {F}^\prime_\lambda({s^\mu},{h}), {F}^\prime_\tau({s^\mu},{h})
 \big] = ({X^\prime}^\rho, h^\prime)\label{Fp0}\ee
 where 
\be {X^\prime}^\rho \equiv  \partial_\lambda s^\mu {\partial x^\beta \over \partial s^\mu}{\partial \over \partial x^\beta}( \partial_\tau s^\rho)-\partial_\tau s^\mu {\partial x^\beta \over \partial s^\mu}{\partial \over \partial x^\beta}( \partial_\lambda s^\rho )\label{vector} \ee
and
\be
h^\prime \equiv (\partial_\lambda s^\mu {\partial x^\beta \over \partial s^\mu}{\partial \over \partial x^\beta} \partial_\tau h - \partial_\tau s^\mu {\partial x^\beta \over \partial s^\mu}{\partial \over \partial x^\beta} \partial_\lambda h ).
\ee

Putting this together we find the invariant action is 
\bea
S_{p=0}&=& \int d^n s \, d\lambda\, d\tau \mid {\partial x^\mu \over \partial s^\nu} \mid  ( {\tilde X}_\alpha {\partial x^\alpha \over \partial s^\rho} +   {\tilde f} {\partial x^\gamma \over \partial s^\rho} {\partial h \over \partial x^\gamma}) \partial_\lambda s^\mu {\partial x^\beta \over \partial s^\mu}{\partial \over \partial x^\beta}( \partial_\tau s^\rho)\cr
&-&\int d^n s \, d\lambda\, d\tau \mid {\partial x^\mu \over \partial s^\nu} \mid( {\tilde X}_\alpha {\partial x^\alpha \over \partial s^\rho} +   {\tilde f} {\partial x^\beta \over \partial s^\rho} {\partial h \over \partial x^\beta})\partial_\tau s^\mu {\partial x^\gamma \over \partial s^\mu}{\partial \over \partial x^\gamma}( \partial_\lambda s^\rho )\cr
&+& \int d^n s \, d\lambda\, d\tau \mid {\partial x^\mu \over \partial s^\nu} \mid {\tilde f}(\partial_\lambda s^\mu {\partial x^\beta \over \partial s^\mu}{\partial \over \partial x^\beta} \partial_\tau h - \partial_\tau s^\mu {\partial x^\beta \over \partial s^\mu}{\partial \over \partial x^\beta} \partial_\lambda h )
\eea
Up to a total $\tau$ derivative, but keeping the total $\lambda$ derivative, we can write
\bea
S_{p=0}&=& \frac{1}{3}\int d^n x \, d\tau  ( {\tilde X}_\alpha +   {\tilde f}  {\partial h \over \partial x^\alpha})\, \partial_\tau s^\mu {\partial x^\alpha \over \partial s^\mu} \cr
&-&\frac{2}{3}\int d^n x \, d\lambda\, d\tau\,  {\tilde f}\,{\partial x^\alpha \over \partial s^\mu}(\partial_\tau s^\mu{\partial \over \partial x^\alpha} \partial_\lambda h-\partial_\lambda s^\mu{\partial \over \partial x^\alpha} \partial_\tau h ) \label{p=0action}
\eea
This action now gives dynamics to the group elements $s(\lambda, \tau, x)$ 
and $h(\lambda, \tau, x)$.   The field ${\tilde X}_\alpha$ and $ {\tilde f}$ are background 
fields that serve as sources for an induced metric $\partial_\tau s^\mu 
{\partial x^\alpha \over \partial s^\mu}$ and a $U(1)$ vector field $\partial h$.   As we will 
explore in the next section, this is a Courant bracket variation of 2D Polyakov gravity along with 
a $U(1)$ WZW model but without central extension.  

\subsection{Review of Kac-Moody and Virasoro Semi-Direct Product \label{pzero}}

We now will make the comparison to this model where 2D Polyakov gravity and WZNW models are combined.  
For completeness let us quickly review the case of the semi-direct product of a Kac-Moody algebra 
with the Virasoro algebra \cite{Lano:1992tc,Lano:1994gx}.  In the mode decomposition the semi-direct product of the two algebras may be written as
\begin{equation}
\left[ L_N,L_M\right]  =(N-M)\,L_{N+M} 
+ c N^3\,\delta _{N+M,0},
\end{equation}
\begin{equation}
\left[ J_N^\alpha ,J_M^\beta \right]  =i\,f^{\alpha \beta \gamma
}\,J_{N+M}^\beta +N\,k\;\delta _{N+M,0}\;\delta ^{\alpha \beta },
\end{equation}
and
\begin{equation}
\left[ L_N,J_M^\alpha \right]  = -M\;J_{N+M}^\alpha, 
\end{equation}
where
$ [\tau^\alpha, \tau^\beta] = i f^{\alpha \beta \gamma} \tau^\gamma. $
One can, for example, realized the algebra on a circle with
\begin{equation}
L_N= \xi_{N}^a \partial_a = ie^{iN\theta }\partial _\theta, \,\,\,\,\,\,\,J_N^\alpha 
= \tau ^\alpha e^{iN\theta }.
\end{equation}
This centrally extended basis can be thought of as the three-tuple,
\begin{equation}
\left(L_A,J_B^\beta ,\rho\right).
\end{equation}
From the above commutation relations, the adjoint representation acts on itself as
\begin{equation}
\left( {L_A,J_B^\beta ,\rho }\right) {*}\left( {L_{N^{\prime }},J_{M^{\prime
}}^{\alpha ^{\prime }},\mu }\right)  =\left( {L_{new},J_{new},\lambda }%
\right) \label{case1-a}
\end{equation}
where
\begin{eqnarray}
L_{new}&=&  \,(A-N^{\prime })\,L_{A+N^{\prime }} \cr
J_{new}&=&  -M^{\prime } J_{A+M^{\prime }}^{\alpha ^{\prime
}}+BJ_{B+N^{\prime }}^{\,\beta }+if\,^{\beta \alpha ^{\prime }\lambda
}J_{B+M^{\prime }}^\lambda\cr
\lambda &=&(c A^3) \delta _{A+N^{\prime },0}+Bk\delta ^{\alpha
^{\prime }\beta }\delta _{B+M^{\prime },0}. \label{case1-b}
\end{eqnarray}

Now lets consider two centrally extended adjoint elements in this algebra, ${\cal F}= (\xi, h, a )$  and ${\cal G}=(\psi, g, b)$ that are functions in this basis.  From the commutation relations, the action of the these adjoint elements on themselves is  
\be
\bigl( \xi, h, a \bigr) \ast \bigl( \psi, g, b\bigr) = 
\biggl( \xi \psi^\prime-\psi \xi^\prime, -\xi g^\prime + \psi h^\prime ,
\int \bigl( {k\over 2 \pi i}  (h'g-g'h)  + {c \over 12 \pi i}  (\xi \psi'''-\psi \xi ''')\bigr) d\theta \biggr),
\ee
where $'$ denotes $\partial_\theta$ the circle parameter and where we have normalized the two-cocycle to be consistent with \cite{Lano:1992tc,Lano:1994gx}. 

Using a closed line integral for the pairing and the two cocycle, the coadjoint action of $\cal{F}$ on a coadjoint vector $\textbf{B}=(D(\theta),A(\theta),\mu)$ may
be written as
\be
\delta_F \textbf{B}   = 
\bigl( -2 \xi'D-D'\xi- i  {c \mu\over 12 \pi} \xi''' -  A {h}', 
 -A'\xi- A\xi' - 2 k i \mu {h}', 0 \bigr). \ee
 
The geometric action for the general Kac-Moody/Virasoro case is
\begin{eqnarray}
&S_{KV}& = \int d\theta\, d\tau \, D(\theta){ \partial_\tau s \over \partial_\theta s} + 
\int d\theta\, d\tau \, d\lambda \, \left(\partial_\tau D~ { \partial_\lambda s \over \partial_\theta s} - \partial_\lambda D~ { \partial_\tau s \over \partial_\theta s}\right) 
\nonumber \\
&-&\int d\theta\, d\tau\, {\rm Tr} A_\theta g^{-1}\partial_\tau g + \int  d\theta\, d\tau \, d\lambda \, {\rm Tr} (\nabla_\lambda A_\theta) g^{-1}
\partial_\tau g - \int   d\theta\, d\tau \, d\lambda \,{\rm Tr}(\nabla_\tau A) g^{-1} \partial_\lambda g 
\nonumber \\
&+& {c \mu  \over 48\pi } \int d\theta\, d\tau \, d\lambda \, \left[
{{\partial^2_{\theta} s}\over{(\partial_{\theta}s)^2}} \partial
_{\tau} \partial_{\theta} s -  {{(\partial^2_{\theta}s)^2
(\partial_{\tau} s)}\over{(\partial_{\theta} s)^3}} \right] d\theta d\tau
     - {k \mu} \int {\rm Tr} g^{-1} {{\partial g}\over{\partial
\theta}}g^{-1}
{{\partial g}\over{\partial \tau}}d \theta d \tau
\nonumber \\
&+& {k \mu} \int {\rm Tr} g^{-1} {{\partial g}\over{\partial \theta}}
\left[ g^{-1}
{{\partial g}\over{\partial \lambda}}, g^{-1} {{\partial g}\over{\partial
\tau}} \right]
d\theta d\tau d\lambda \label{kacvirasoro}
\end{eqnarray}
where the derivative operators are defined by
\be
\nabla_\tau \equiv \partial_\tau - 
{\partial_\tau s \over \partial_\theta s} \partial_\theta 
- \partial_\theta ( {\partial_\tau s \over \partial_\theta s}), \qquad 
\nabla_\lambda \equiv \partial_\lambda - {\partial_\lambda s \over \partial_\theta s} 
\partial_\theta - \partial_\theta ( {\partial_\lambda s \over \partial_\theta s}).
\ee
This action will give intuition to the action we derive from the Courant bracket when $p=0$.

\subsection{Central Extension of the $p=0$ Courant Bracket in One Dimension}

Now let us examine the issue of central extensions for the one dimensional $p=0$ Courant 
bracket when there are periodic boundary conditions or when the vector fields vanish on the 
boundary.  We define a centrally extended Courant bracket for $p=0$ by writing the algebra of a triplet as;  
\be
\eqlabel{courant1}
[(X,f,\textsc{c}),(Y,g,\textsc{d})]=\left([X,Y], \,\call_X g-\call_Y f\,,{\cal C}_1(X,Y)+{\cal C}_2(f,g) \right).
\ee
Here \textsc{c} and \textsc{d} are centrally extended elements of the Bracket.
The two cocycles ${\cal C}_1(X,Y)$ and ${\cal C}_2(f,g)$  are defined as 
\begin{equation}{\cal C}_1(X,Y) = \beta_1 \int (X^a \nabla_a \nabla_b 
\nabla_c  Y^c)\,dx^b,
\end{equation}
and
\begin{equation} 
{\cal C}_2(f,g) = \beta_2 \int \left(f (\nabla_b g )  -g (\nabla_b f )\right)\,dx^b.
\end{equation}
These two-cocycles vanish under the Jacobi Identity, 
\bea
\hskip10pt [(X,a,\textsc{a}),[(Y,b,\textsc{b}),(Z,c,\textsc{c})]]&+&\cr [(Z,c,\textsc{c}),[(X,a,\textsc{a}),(Y,b,\textsc{b})]]&+&
[(Y,b,\textsc{b}),[(Z,c,\textsc{c}),(X,a,\textsc{a})]]=0.
\eea
This follows since
\be
{\cal C}_1(X,[Y,Z])+{\cal C}_1(Y,[Z,X])+{\cal C}_1(Z,[X,Y])
=0
\ee
and since
\bea
&&{\cal C}_2\bigg(a,\call_Y c - \call_z b\bigg)+{\cal C}_2\bigg(b,\call_Z a - \call_X c\bigg)+{\cal C}_2\bigg(c,\call_X b - \call_Y a\bigg) \cr
&&= \int a(Y c^\prime-Z b^\prime)^\prime\,dx -\int a^\prime (Y c^\prime -Z b^\prime)\,dx+ \int b(Z a^\prime -X c^\prime)^\prime \,dx\cr
&&-\int b^\prime (Z a^\prime -X c^\prime) \,dx+ \int c(X b^\prime -Y a^\prime)^\prime\,dx - \int c^\prime (X b^\prime - Y a^\prime)\,dx=0\cr&&
\eea
when there are periodic boundary conditions or when the vector fields vanish at the spatial boundaries.  
In order to maintain a manifestly covariant description of the coadjoint representation, we us 
a pairing akin to Eq.[\ref{pairing1}] viz.,
\be  
\langle \textbf{B} \mid {\cal F} \rangle = \langle (D_{\rho \gamma},A_\rho, \tilde{\beta}) 
\mid (X^\beta, f, \textsc{a}) \rangle = \int X^\beta D_{\beta \gamma} dx^\gamma + \int A_\rho\, f \, 
dx^\rho + \tilde{\beta} \textsc{a}.
\ee
We find that the action of the adjoint element $\cal{F}$ on  a coadjoint element
$\textbf{B}=(D(\theta), A(\theta), {\tilde \beta})$ may
be written as
\be
\delta_F \textbf{B}   = 
-\bigl( 2 X^\prime D+D^\prime X +  {\tilde \beta}  X^{'''} + A {f\,}^\prime, 
 A^\prime X + A X^\prime + 2 {\tilde \beta} {f\,}^\prime, 0 \bigr). \ee
The geometric action can be read off and is 
\begin{eqnarray}
&&\,\,\,\,\,\,\,\ S_{p=0,d=1} = \cr
&&\int d\lambda\,d\tau\,ds^\gamma \,\biggl(D_{\alpha \beta}(s(x)) + A_\alpha \partial_\beta f + 
(\partial_\mu \Gamma^\mu_{\alpha \beta} -\frac{1}{2} \Gamma^\mu_{\alpha \nu} \Gamma^\nu_{\beta \mu}) \,\biggr)\frac{dx^\alpha}{ds^\rho}\frac{dx^\beta}{ds^\gamma}\bigl( \frac{\partial s^\xi}{\partial \tau} \frac{\partial^2 s^\rho}{\partial s^\xi \partial \lambda} - \frac{\partial s^\xi}{\partial \lambda} \frac{\partial^2 s^\rho}{\partial s^\xi \partial \tau}\bigr)\cr
&\,& +\int d^\beta x \, d\lambda\, d\tau\, \bigl( A_\beta + 2 {\tilde \beta}\, \partial_\alpha f \bigr)\,{\partial x^\alpha \over \partial s^\mu}(\partial_\tau s^\mu{\partial \over \partial x^\alpha} \partial_\lambda f -\partial_\lambda s^\mu{\partial \over \partial x^\alpha} \partial_\tau f ). 
\end{eqnarray}
To compare let $g$ be the $U(1)$ group element such that $g=\exp(i f)$.
Then after integrating by parts and suppressing the one dimensional indices we find  
\begin{eqnarray}
&S_{p=0,d=1}& = \int d^2x\, (D - i A g^{-1}\partial_x g) { \partial_\tau s \over \partial_x s} \cr 
&+&
\int d^3 x \left(\partial_\tau (D -  i A g^{-1}\partial_x g) ~ { \partial_\lambda s \over \partial_x s} 
- \partial_\lambda (D - i A g^{-1}\partial_x g) ~ { \partial_\tau s \over \partial_x s}\right) 
\nonumber \\
&+&i \int d^2x  A_x g^{-1}\partial_\tau g - i \int d^3x  (\nabla_\lambda A_x) g^{-1}
\partial_\tau g + i \int d^3x (\nabla_\tau A) g^{-1} \partial_\lambda g 
\cr
&+& {1 \over 4} \int d^2x\,  \left[
{{\partial^2_{x} s}\over{(\partial_{x}s)^2}} \partial
_{\tau} \partial_{x} s -  {{(\partial^2_{x}s)^2
(\partial_{\tau} s)}\over{(\partial_{x} s)^3}} \right] \cr
     &+& i \,{\tilde \beta \over 2} \int d^2x\, g^{-1} {{\partial g}\over{\partial
x}}g^{-1}
{{\partial g}\over{\partial \tau}} 
\end{eqnarray}
This is to be compared with Eq.[\ref{kacvirasoro}] when the group of gauge transformations 
is Abelian.  This now makes the U(1) Kac-Moody/Virasoro analogue that we had mentioned earlier complete.

\subsection{A geometric action for the Courant bracket ($p \ne 0$)\label{explicitall}}
The general case $p \ne 0$ has a number of new features. This can be traced to the fact that 
the Courant bracket does not satisfy the Jacobi identity for general $p\ne 0$, except in a restricted sense 
as explained in \cite{gualtieri}.  Our purpose is to construct a geometric action 
built from a symplectic two-form $\Omega$.  This requires that $\Omega$ be non-degenerate 
and closed, i.e. $d \Omega =0$.  For the geometric action the closed condition, i.e. $d\Omega=0$, arises as 
a consequence of the Jacobi identity being satisfied by the defining algebra.  Therefore in this section we will be required to restrict our attention to the case where the Jacobi identity is satisfied.

A sufficient condition for a subbundle of $T\oplus \wedge^p T^*$ to satisfy the Jacobi identity is that the vectors fields should be hypersurface orthogonal to the p-forms. Then for all elements of this subbundle, $(X, \xi)$, we require that $X^{a_1} \xi_{a_1 a_2 \cdots a_p}=0$.  
From Eq.[\ref{deltaX}],
we can identify the group action on $(\tilde X_a,\tilde \xi^{a_1 \ldots a_p})$.   Note the 
exterior derivative terms in the Eq.[\ref{deltaX}] will not contribute because of the hypersurface orthogonality condition. 
It is straightforward to write the coadjoint element analogous to the Eq.[\ref{Bp0}], 
\be
\textbf{B}^\prime({s^m},{\xi}) = \bigg(\mid {\partial x^m \over \partial s^n} \mid   {\tilde X}_\alpha {\partial x^a \over \partial s^r} , \,\, \mid {\partial x^m \over \partial s^n} \mid {\tilde \xi}^{b_1 \ldots b_p} \bigg)\label{Bpp}\ee

The adjoint element corresponding to Eq.[\ref{Fp0}] is given by 
\be
\big[ \textbf{F}^\prime_\lambda({s^m},{\epsilon}), \textbf{F}^\prime_\tau ({s^m},{\epsilon})
 \big] = ({X^\prime}^a, \epsilon^\prime_{a_1 \ldots a_p})\label{Fpp}\ee
 where 
\be {X^\prime}^a \equiv  \partial_\lambda s^m {\partial x^b \over \partial s^m}{\partial \over \partial x^b}( \partial_\tau s^a)-\partial_\tau  s^m {\partial x^b \over \partial s^m}{\partial \over \partial x^b}( \partial_\lambda s^a )\label{vectorp} \ee
and a $p$-form $\b_{a_1 \ldots a_p}$ which is in the Dirac subbundle.

As before we can put this all together and find that,
\bea
&&S_{p}= \int d^n s \, d\lambda\, d\tau \mid {\partial x^m \over \partial s^n} \mid  ( {\tilde X}_d {\partial x^d \over \partial s^a} \left(\partial_\lambda s^m {\partial x^b \over \partial s^m}{\partial \over \partial x^b}( \partial_\tau s^a)-\partial_\tau  s^m {\partial x^b \over \partial s^m}{\partial \over \partial x^b}( \partial_\lambda s^a )\right)\cr
&+& \int d^n s \, d\lambda\, d\tau \mid {\partial x^m \over \partial s^n} \mid {\tilde \xi}^{b_1 \ldots b_p} \b_{b_1 \ldots b_p}\cr \label{Paction}
\eea
The fields ${\tilde X}_d$ and ${\tilde \xi}^{b_1 \ldots b_p}$ serve as background fields that dictate together the symmetry of the action. 

As an explicit example we may consider a $p+1$-form $B$ and define,
\be \b^\prime_{a_1 \ldots a_p} \equiv B_{b_1 \ldots b_p b_{p+1}}  {\partial s^{b_1} \over \partial x^{a_1}} \ldots {\partial s^{b_p} \over \partial x^{a_p}}{X^\prime}^{b_{p+1}}.\ee This corresponds to a $B$-field transformation of the zero $p$-form  with the automorphism discussed earlier, Eq.[\ref{automorphism}].
The physics of membranes offers an interpretation of the above action.  A $p$-brane is charged by a $p+1$-form via the following interaction,
\be
S_p = -\int d\,\tau \, d\sigma^1 \cdots d\sigma^p\,  {\partial s^{\mu_1} \over \partial \sigma^1} \cdots {\partial s^{\mu_p} \over \partial \sigma^p}\,\left( {\partial s^{\mu} \over \partial \tau}{B}_{\mu \mu_1 \ldots \mu_p} \right). 
\ee This action measure the flux through the $p$ dimensional spatial surface.  
Furthermore the $p+1$-form has it dynamics through
\be
S_{\tilde \xi}= \int d^nx \, H_{\m_1 \cdots \m_{p+2}} \,H^{\m_1 \cdots \m_{p+2}},
\ee
where 
\be H_{\m_1 \cdots \m_{p+2}} \equiv \partial_{[\m_1} {B}_{\mu_2 \ldots \mu_{p+2}]}. \label{Haction}
\ee
When $n=p$ in Eq.[\ref{Paction}], one can interpret the field ${\beta}_{b_1 \ldots b_p}$ as charging a one parameter family ($\l$) of p-branes embeddings given by ${\tilde \xi}^{{b_1 \ldots b_p}}$.  The $2^{nd}$ summand of Eq.[\ref{Paction}] corresponds to the response of the flux to this family of embeddings as $\l$ goes from $0$ to $1$, where at say $\l=0$ the map is the identity and at $\l=1$ the map is a particular $p$-brane embedding.  

We are interested in how this action may be used to follow changes in sympletic and complex structures that are related to the Courant algebroid on $T\oplus \wedge^p T^*$.    Since each orbit has a distinct symplectic structure, one now has a method of characterizing the symplectic structures in terms of the orbits and embedding manifold.  By incorporating the Lagrangian dynamics for the ${\tilde X}_d$ and ${\tilde \xi}^{b_1 \ldots b_p}$ fields one effectively has a theory of symplectic structures that perhaps can be related to generalized complex structures and geometric quantization. In future work we will use the methods of \cite{Branson:1998bc}, to determine the dynamics of these fields which is transverse to the orbits, i.e. the {\em transverse action}.  Thus the field theory of ${\tilde X}$ and ${\tilde \xi}$ provides an effective potential for a family of symplectic structures.  These symplectic structures might be related to generalized complex structures which in turn would give an effective action for the space of generalized complex structures.  This would give a variational principle to the complex structures which can assist in studying the string landscape problem.   

\section{Extended Complex and K\"ahler Structures \label{6}}
In this section we give several examples on how one might employ $\Omega_{\tilde {\cal Z}}$ to extend both K\"ahler geometry and generalized complex structures to the space of orbits.  At present we have not made contact with  sigma models and supersymmetry but do hope to address these relationships in future work.  
\subsection{A K\"ahler Structure on the $p=1$ Orbits of ${\tilde {\cal Z}}$}
A natural question to ask is if one can endow the coadjoint orbit of ${\tilde Z}=(Z_a, \zeta^a)$ with a K\"ahler geometry.  Suppose that $\d{\tilde A}$ and $\d{\tilde  B}$ are on the obit of ${\tilde Z}$ where $\d{\tilde A} = \d_{\cal X}\,{\tilde {\cal Z}}$ and a similar relation for ${\cal Y}$ and $\d{\tilde  B}$.  Can one define a K\"ahler-like metric on the orbit?  Assume that a suitable almost complex structure exist on the orbit of ${\tilde Z}$.  Then one can write a general K\"ahler metric ${\cal G}_{\tilde Z}(\d{\tilde A},\d{\tilde B}) \equiv <<{\cal X},{\cal Y}>>_{\tilde Z}$ 
as
\be 
<<{\cal X,Y}>>_{{\tilde {\cal Z}}}=\frac{1}{2}\left(- \Omega_{{\tilde {\cal Z}}}({\cal J X, Y}) + \Omega_{{\tilde {\cal Z}}}({\cal X, J Y}) \right). 
\ee
To be explicit we write ${\cal X} = (X^a, \xi_a)$ and ${\cal Y}=(Y^a, \eta_b)$.
Consider the usual almost complex structure with Riemannian metric on $T\oplus T^*$ given by
\be {\cal J}_g = \left(
\begin{array}[h]{cc}
0 & -g^{ ab}\\
g_{a b} & 0	
\end{array}
\right)
\ee
so that ${\cal J}({\cal X})=(-\xi^a, \, X_a)$ and a similar action on ${\cal Y}$. 
Then a direct calculation shows that on the orbit of ${\tilde {\cal Z}}$ we have a generalized K\"ahler metric given by
\bea
<<{\cal X},{\cal Y}>> &=& \frac{\alpha_1}{2} \int {\tilde z}_a \left( - {\cal L}_X \eta^a -{\cal L}_Y \xi^a \right) \nonumber \\
&+& \frac{\alpha_2}{2} \int {\tilde \zeta}^b \left( {\cal L}_Y X_b + {\cal L}_X Y_b \right) \nonumber \\
&+& \frac{\alpha_2}{2} \int {\tilde \zeta}^b \left({\cal L}_{\xi} \eta_b + {\cal L}_{\eta} \xi_b \right).
\eea

\subsection{Generalized Complexification using $\Omega_{\tilde Z}$}
As one can see in Eq.[\ref{twoform}], $\Omega_{\tilde Z}$ will not map elements of $T$ into $T^*$.  
However part of our motivation for this work was to exploit the rich symplectic geometry of the coadjoint orbits and extend it to generalized complex structures for the case where $p=1$, i.e. $T\oplus T^*$, and also to see whether this method can lead to $p$-extended complexifications.  The space of orbits would then foliate the different complexifications.  In this section we wish to find an analogue of the almost complex structure of the type
\be {\cal J}_\omega = \left(
\begin{array}[h]{cc}
0 & -\omega^{-1}\\
\omega & 0	
\end{array}
\right)
\ee discussed in \cite{hitchin,hitchinkitp,gualtieri}
with $\Omega_{\tilde Z}$ replacing $\omega$.  
Now $\Omega_{\tilde Z}({\cal X},{\cal Y})$ maps $T\oplus \wedge^p T^*$ into $T^* \oplus \wedge^p T$.  Therefore we look for a complexification on the space ${\cal E}_p \in  (T\oplus \wedge^p T^*) \otimes (T^* \oplus \wedge^p T$).  In what follows we assume there exists a Riemannian metric that maps $T$ into $T^*$ and vice versa.
Then, for two coordinates ${\bf X}=({\cal X, Y^*})$ and ${\bf Y}=({\cal W, Z^*})$ on ${\cal E}_p$ we define the  bracket $[[*,*]]$ as 
\be 
[[ {\bf X},{\bf Y}]] = {\cal [X,W]}\otimes {\cal [(Y^*)^*, (Z^*)^*]^*},
\ee
with $[*,*]$ the Courant bracket. 
Then for the phase space coordinate, ${\bf X}=({\cal X},{\cal Y}^*)$ we define the almost complex structure as 
\be  {\cal J}_{\tilde Z}({\bf X})={\cal J}_{\tilde Z}({\cal X},{\cal Y}^*) = \left(-(\Omega_{\tilde Z})^{-1}({\cal Y}^*), \Omega_{\tilde Z}({\cal X})\right).\ee
This almost complex structure extends to all $p$ and the isotropies of ${\tilde Z}$ determine the symmetries of this $p$-extended almost complex structure. 
 
For the $p=1$ case ${\cal E}_1$ is just two copies of $T\oplus T^*$. Furthermore we can use the components of $\Omega_{\tilde Z}$, viz.  $\o(1), \o(2),$ and $\o(3)$ in Eqs.[\ref{p1omega}] to write a complex structure directly on $E=T\oplus T^*$.  Since $\o(1)$ is closed and invertible independent of $\o(2)$ and $\o(3)$.  We can write a complex structure associated with the orbit of ${\tilde Z}$ as
\be
{\cal J}_{Z}=\left( \begin{array}[h]{cc}
0 & -{\omega(1)}^{-1} \\
{\omega(1)} &0 \end{array}\right).\label{ComplexP1}
\ee
In a similar vein, since $\O^{-1}_{\tilde Z}$ corresponds to the Poisson bracket algebra, and $\o(2)$ and $\o(3)$ correspond to linear transformations on the Courant elements, we can write a different almost complex structure as 
 \be
{\hat{\cal J}}_{Z}=\left( \begin{array}[h]{cc}
0 & -{\omega(3)}\,{\omega(1)}^{-1} \,{\omega(2)}  \\
{\omega(2)}^{-1}\,\omega(1) \,{\omega(3)}^{-1} &0 \end{array}\right).\label{ComplexP2}
\ee

\subsection{$p$-Extended K\"ahler Structure on $p+1$ Dimensional Submanifolds} 
 Eq.[\ref{Omega}] defines a symplectic structure on the orbits associated with Courant algebroid  $T\oplus \wedge^p T^*$ that has a Dirac structure.  In this section we give an example for an extension of generalized complex structures to $T\oplus T^*$ by using the Levi-Civita tensor on a $p+1$ submanifold.   This interest is partly due to $D$-brane physics as briefly stated above.
Motivated by the Hodge-$*$ duality between one forms and $p$-forms on $p+1$ dimensional submanifolds we define an almost complex structure ${\cal J}((X,\xi))$ on $T\oplus \wedge^p T^*$ with basis $T\otimes \wedge^p T^*$ by writing
\be
{\cal J}\left({\cal X}\right)={\cal J}\left((X^b, \eta_{\,b_1 \cdots b_p})\right)=\left(-\frac{1}{\sqrt{p!}} \epsilon^{\,b\, b_1 \cdots b_p} \eta_{\,b_1 \cdots b_p}, -\frac{1}{\sqrt{p!}} \epsilon_{\,b\, b_1 \cdots b_p} X^b \right),
\ee
where $\epsilon(\sigma)_{b_1 \cdots b_{p+1}}$ is a $p+1$ dimensional Levi-Civita tensor on a submanifold say $\sigma$. 
Now in general, the Nijenhuis tensor associated with ${\cal J}$ can be computed from 
\be
N_J(\cal{X},\cal{Y}) = [J(\cal{X}),J(\cal{Y})]-J([J(\cal{X}),\cal{Y}])+J([J(\cal{Y}),\cal{X}])-[\cal{X},\cal{Y}].
\ee
In tensor notation this becomes 
\be
N_J({\cal X,Y})=(Z_J({\cal X,Y})^b, \zeta_J({\cal X, Y})_{\, b_1 \cdots b_p}),
\ee
with
\be
Z_J({\cal X,Y})^b = 0,
\ee
and
\bea
\zeta_J({\cal X, Y})_{\, a_1 \cdots a_p}&=& \partial_{a_1}(Y^m  \xi_{\, m a_2 \cdots a_p} + \cdots + Y^m \xi_{\, a_1 \cdots a_{p-1} m}) \nonumber \\
&-& \partial_{a_1}( X^m  \eta_{\,m a_2 \cdots a_p} + \cdots + X^m \eta_{\, a_1 \cdots a_{p-1} m}). \nonumber \label{complex2}
\eea
As one can see the obstruction for the vanishing of the Nijenhuis tensor is removed by the conditions of the Dirac subbundle. 

From this complex structure and the symplectic two-form $\Omega$, we can define the $p$-extended generalized K\"ahler metric, $<<*,*>>$, for the Courant algebroid with Dirac structure on the $p+1$  dimensional submanifold $\sigma$ on each orbit through,
\be 
<<{\cal X,Y}>>_{\bf ({\tilde {\cal Z}},\sigma)}=\frac{1}{2}\left(- \Omega_{\bf ({\tilde {\cal Z}},\sigma)}({\cal J X, Y}) + \Omega_{\bf ({\tilde {\cal Z}},\sigma)}({\cal X, J Y}) \right). 
\ee
Here ${\tilde {\cal Z}}=(\tilde{z_a}, \tilde{\zeta}^{b_1 \cdots b_p})$ is the dual element defining the orbit, while 
${\cal X}=(x^a, \xi_{a_1 \cdots a_p})$ and ${\cal Y}=(y^a, \eta_{a_1 \cdots a_p})$ are elements of the Dirac algebroid. 
With this we may write the metric explicitly as 
\bea
<<{\cal X,Y}>>_{\bf ({\tilde {\cal Z}},\sigma)}&=&\frac{\alpha_1}{2} \int_{\sigma} \left( {\cal L}_y \xi^a + {\cal L}_x \eta^a \right) {\tilde{z_a}} 
-\frac{\alpha_2}{2} \int_{\sigma} \left( \xi^a \nabla {\tilde \eta} + \eta^a \nabla_a {\tilde \xi} \right) \nonumber \\
&+& \frac{\a_2}{2} \int_{\sigma} (\nabla_a {\tilde \zeta}^{a_1 \cdots \a_p}) \{ \xi^a \eta_{a_1 \cdots a_p} +\eta^a \xi_{a_1 \cdots a_p} \} \nonumber \\
&-& \frac{ p \a_2}{2}  \int_\sigma \left({\tilde \eta}_a^{\, b} \nabla_b \xi + {\tilde \xi}_a^{\,b} \nabla_b \eta^a - {\tilde \zeta}^b_{\, a c}(x^c \nabla_b y^a + y^c \nabla_b x^a)     \right) \nonumber\\
&+& \frac{\alpha_2}{2} \int_\sigma {\tilde \zeta}_b \left(x^a \nabla_a x^b + x^a \nabla_a y^b\right),
\eea
where in the above 
\bea
&&{ \eta}^b \equiv \frac{1}{\sqrt{p\!}} \,{\eta}_{a_1 \cdots a_p} \epsilon^{b\, a_1 \cdots a_p}  \hskip.5in { \xi}^b \equiv \frac{1}{\sqrt{p\!}} \,{\xi}_{a_1 \cdots a_p} \epsilon^{b\, a_1 \cdots a_p} \nonumber\\
&&{\tilde \zeta}_b \equiv \frac{1}{\sqrt{p\!}} {\tilde \zeta}^{a_1 \cdots a_p} \epsilon_{b\, a_1 \cdots a_p} \hskip.5in 
{\tilde \zeta}^{a_1\,}_{\,\,\,\,\, a c}\equiv {\tilde \zeta}^{a_1 \cdots a_p}  \epsilon_{a a_2 \cdots a_p \,c}\nonumber\\
&&{\tilde \eta}\equiv {\tilde \zeta}^{a_1 \cdots a_p} \eta_{a_1 \cdots a_p} \hskip.9in{\tilde \xi}\equiv {\tilde \zeta}^{a_1 \cdots a_p} \xi_{a_1 \cdots a_p} \nonumber\\
&&{\tilde \eta}_{a\,}^{\,a_p} \equiv {\tilde \zeta}^{a_1 \cdots a_p} \eta_{a_1 \cdots a_{p-1} a}  \hskip.5in  {\tilde \xi}_{a\,}^{\,a_p} \equiv {\tilde \zeta}^{a_1 \cdots a_p} \xi_{a_1 \cdots a_{p-1} a}.  \nonumber \\
\eea
\section{Conclusions}\label{conclusions}
In this paper we have constructed geometric actions based on the Kirillov-Kostant symplectic  
form, for the Courant bracket. The orbit method provides an interesting starting point for studying the changes in certain generalized complex structure by assigning a complex structure (through the symplectic geometry) to an orbit. Changing orbits is tantamount to changing the equivalence class of generalized complex structures.   In this note we work out the details and offer interpretations of the geometric actions associated with Dirac structures on $T\oplus \wedge^p T^*$.   Our first result is given in subsection \ref{pzero} which contains 
a geometric action for the centrally extended Courant bracket in the case of $p=0$ in one spatial dimension. This action is a generalization of Polyakov's 2-d quantum gravity. Its structure is similar to the geometric action 
for the semidirect product of Virasoro and the affine Kac-Moody algebra with group $U(1)$. The general case 
$p\ne 0$ is restricted since closure of the Kirillov-Kostant two form is guaranteed only by the Jacobi 
identity on the Courant bracket. This last property requires  pairs $(X,\xi)$ to form  
a Dirac structure. More concretely, the vector $X$ and the form $\xi$ must be orthogonal in the sense of 
interior product, $i_X \xi=0$.  We were then able to construct the geometric action for all $p$ restricted to the Dirac structure.  The action lends itself to an interpretation of integrated fluxes due to similarities of the geometric action  with charged $p$-branes.  We next considered K\"ahler geometries on the orbits for $p=1$, and some unique almost complex structures related to the symplectic two-form on the orbit.  This gives us a way of treating the orbit as a complex geometric space that is suitable.  We then consider $p$-extensions of generalized complex structures using two examples.  One corresponded to doubling the space to include the dual of the Courant algebroid (two field theoretic dimensions) and the other relies on the Hodge * dual to relate $p$ forms to vector fields on a $p+1$ dimensional manifold.  In that case we again went further to show a generalization of K\"ahler geometry on the orbits of $E_p$.  

The above work has been restricted to a Dirac structure, 
and the question remains concerning the direct connection to the $B$-field in string theory. This needs to be further studied but the method can be adapted to the full $T\oplus \wedge^p T^*$ since in the general  case the Jacobiator will differ from zero only by the exterior derivative of a Nijenhaus tensor \cite{gualtieri}.  Although we have not shown this, we suspect that this addition will add to the geometric action dynamical terms such as Eq.[\ref{Haction}].  Lifting the Dirac structure may bring this work closer to the physics the $B$ field case in $p=1$ and more generally the study of D-brane fluxes.  Part of our future work will try to make these remarks concrete. We also intend to make stronger the relationship with the extensive work of generalized complex structure by incorporating a spinor field associated with each orbit and relating this to supersymmetry.  

There are many open problems that remain.  For example, having a generalization 
of Polyakov 2-d quantum gravity one wonders about the quantization aspects of such action and its 
possible interpretation as a generalization of string theory. It would also be interesting to connect 
the action obtained in our paper with other approaches based on the 
Batalin-Vilkovisky formalism recently discussed in \cite{zuc}. There are similarly interesting works motivating 
actions based on current algebras for Courant brackets, and that, as in our case, make crucial use of the Dirac structure \cite{Alekseev:2004np} and \cite{Bonelli:2005ti} Another question is what is the relationship of the different $p$ on $n$ dimensional manifolds and the homology classes of these manifolds.  Such a relationship could bring this work closer to understanding the moduli of Calabi-Yau compactifications.  We see this work as an initial probe into the study of inherent structures of the Courant bracket in the investigations of generalized complex and K\"ahler structures.  

\section*{Acknowledgments}
We are grateful to B. Uribe for various enlightening conversations on generalized complex geometry and 
the Courant bracket in particular. This work is  partially supported by Department of Energy under
grant DE-FG02-95ER40899 to the University of Michigan and by a grant from the National Science Foundation PHY 02-44377 awarded to the University of Iowa.


\end{document}